\documentclass[preprint2]{aastex}

\defcitealias{Kalvans15b}{Paper~II}
\defcitealias{Kalvans15a}{Paper~I}
\slugcomment{To appear in The Astrophysical Journal}

\shorttitle{Ices in starless cores}
\shortauthors{Kalv\=ans}

\begin{document}


\title{Ice chemistry in starless molecular cores}

\author{J. Kalv\=ans}
\affil{Engineering Research Institute "Ventspils International Radio Astronomy Center" of Ventspils University College,
    Inzenieru 101, Ventspils, Latvia, LV-3601}
\email{juris.kalvans@venta.lv}

\begin{abstract}
Starless molecular cores are natural laboratories for interstellar molecular chemistry research. The chemistry of ices in such objects was investigated with a three-phase (gas, surface, and mantle) model. We considered the center part of five starless cores, with their physical conditions derived from observations. The ice chemistry of oxygen, nitrogen, sulfur, and complex organic molecules (COMs) was analyzed. We found that an ice-depth dimension, measured, e.g., in monolayers, is essential for modeling of chemistry in interstellar ices. Particularly, the H$_2$O:CO:CO$_2$:N$_2$:NH$_3$ ice abundance ratio regulates the production and destruction of minor species. It is suggested that photodesorption during core collapse period is responsible for high abundance of interstellar H$_2$O$_2$ and O$_2$H, and other species synthesized on the surface. The calculated abundances of COMs in ice were compared to observed gas-phase values. Smaller activation barriers for CO and H$_2$CO hydrogenation may help explain the production of a number of COMs. The observed abundance of methyl formate HCOOCH$_3$ could be reproduced with a 1kyr, 20K temperature spike. Possible desorption mechanisms, relevant for COMs, are gas turbulence (ice exposure to interstellar photons) or a weak shock within the cloud core (grain collisions). To reproduce the observed COM abundances with the present 0D model, 1-10\% of ice mass needs to be sublimated. We estimate that the lifetime for starless cores likely does not exceed 1Myr. Taurus cores are likely to be younger than their counterparts in most other clouds.
\end{abstract}


\section{Introduction}
\label{intro}

Low-mass interstellar molecular clouds, such as the Bok globules, are a model case for cloud structure and star-formation studies, thanks to their relative simplicity and proximity \citep{Hartmann09}. For astrochemistry, stable starless cores within these clouds are particularly interesting. They are sufficiently dense for a significant mass of ice to accumulate onto the grains, supposedly not disturbed by gravitational contraction motions. Thus, modeling of surface chemistry can be simplified to calculations that consider constant, yet realistic physical conditions.

The aim of the present paper is to study the evolution of the chemical composition of ices in starless cores with physical conditions that are relevant to actual cores. The modeling of surface chemistry in starless cores has a long history. Important papers that have presented significant advances in the field include \citet{Watson72,Allen77,Pickles77a,Pickles77b,Tielens82,Hendecourt85,Brown89,Brown90,Hasegawa92}. A model for small, diffuse clouds (not the starless core) with surface chemistry has been presented by \citet{Turner98}. Such two-phase models with constant physical parameters have also been used in recent papers \citep[e.g.,][]{Du12,Vasyunin13b}.

Starless cores with physical conditions derived from actual observations have also been modeled \citep{Du12,Maret13,Lippok13,Awad14}. Recent advances in the field of interstellar ice chemistry in combination with available observational data \citep[e.g.][]{Launhardt10} make the modeling of ices in quiescent cores an interesting proposal. These advances include ices described as consisting of layers, with only the surface monolayer (ML) being avalaible for reactions \citep{Hasegawa93b,Bergin97,Cuppen07,Garrod11,Taquet12}, photoprocessing of subsurface ices \citep{Kalvans10,Garrod13a,Chang14}, and a realistic model of ice molecule formation.

An important aspect in interstellar ice models is a proper description of surface reactions in a \textit{contracting} molecular core. Notably, this includes the surface synthesis of CO$_2$ \citep{Ruffle01,Garrod11}. \citet{Brown88} and \citet{Aikawa01} show that the contraction phase is particularly important for a correct representation of the ice formation epoch. Recent dense core models that focus on surface chemistry often do not consider the core contraction phase \citep{Du12,Kalvans13,Vasyunin13b,Chang14,Reboussin14}. The dependence of ice composition on the core density and interstellar extinction has been shown by a number of earlier studies \citep[e.g.,][]{Rawlings92,Viti99,Green01}

We studied four starless cores -- CB~17~-~SMM1, CB~26~-~SMM, CB~27~-~SMM, and B~68 -- with chemical kinetics modeling. They were treated as clumps of gas and dust with an increased density towards the center, surrounded by a more diffuse cloud. The cores undergo an initial contraction phase, until they reach a state with constant (observed) physical conditions (section~\ref{phys}). A detailed surface and subsurface ice chemistry model has been used (\ref{chem}), whose application to starless cores is the main novelty of this paper.

\section{Methods}
\label{meth}

The model `Alchemic-Venta' was employed in this research, taken from previous works \citep[][hereafter Papers I and II, respectively]{Kalvans15a,Kalvans15b}. It is based on the `ALCHEMIC' code \citep{Semenov10}, with details explained below. Physical parameters in the model were adjusted for a selection of non-starforming cores. Papers I and II investigate collapsing prestellar cores, while the present paper focuses on stable, non-starforming cores.

\subsection{Physical model}
\label{phys}

\subsubsection{Cloud core conditions}
\label{ccond}
\begin{table*}
\begin{center}
\footnotesize
\caption{The constant physical parameters of modeled starless cores during their stable Phase~2.}
\label{tab-phys}
\begin{tabular}{r|cccccc|ccc}
\tableline\tableline
& \multicolumn{6}{c}{Observational data} & \multicolumn{3}{c}{Calculated data} \\
Core & $n_{\rm H,0}$, cm$^{-3}$ & $r_0$, AU & $r_1$, AU & $\eta$ & $N_{\rm out}$, cm$^{-2}$ & Ref. & $A_V$, mag & $T$, K & $t_1$, kyr \\
\tableline
CB 17 S - SMM1  & 2.3E+5 & 9.5E+3 & 3.0E+4 & 4.9 & 0 & S\tablenotemark{a} & 13.8 & 8.2 & 1465 \\
CB 17 L - SMM & 1.3E+5 & 1.1E+4 & 3.0E+4 & 5.0 & 3.6E+20 & L\tablenotemark{b} & 9.1 & 9.2 & 1451 \\
CB 26 - SMM & 8.7E+4 & 8.0E+3 & 4.0E+4 & 3.0 & 2.2E+20 & L & 6.5 & 11.1 & 1437 \\
CB 27 - SMM & 1.3E+5 & 1.3E+4 & 6.6E+4 & 6.0 & 3.1E+20 & L & 9.5 & 8.9 & 1451 \\
B 68 - SMM & 4.0E+5 & 7.0E+3 & 2.7E+4 & 5.0 & 3.0E+20 & L & 17.6 & 7.9 & 1482 \\
\tableline
\end{tabular}
\tablecomments{Assumed physical parameters: $n_{\rm H,0}$--density at the center of the core; $r_0$--radius of the central density plateau; $\eta$--power-law slope for density; $r_1$--radius of the region (core) in consideration; $N_{\rm out}$--column density of the surrounding cloud. Calculated parameter $t_1$ is the length of the cloud core collapse Phase~1.}
\tablenotetext{a}{\citet{Schmalzl14}}
\tablenotetext{b}{\citet{Lippok13}}
\end{center}
\end{table*}
%
\begin{table}
\begin{center}
\footnotesize
\caption{Initial abundances of chemical species.}
\label{tab-ab}
\begin{tabular}{rc}
\tableline\tableline
Species & $n_i/n_{\rm H}$ \\
\tableline
H$_2$ & 0.4995 \\
H & 1.00E-03 \\
He & 9.00E-02 \\
C & 1.40E-04 \\
N & 7.50E-05 \\
O & 3.20E-04 \\
Na & 2.00E-08 \\
Mg & 2.55E-06 \\
Si & 1.95E-06 \\
P & 7.63E-08 \\
S & 1.50E-06 \\
Cl & 1.40E-08 \\
Fe & 7.40E-07 \\
\tableline
\end{tabular}
\end{center}
\end{table}
\begin{figure*}
 \vspace{-3cm}
  \includegraphics[width=18.0cm]{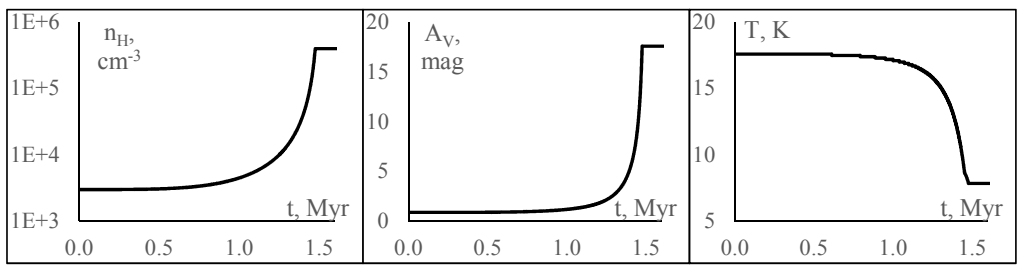}
 \vspace{-18cm}
 \caption{Density of hydrogen atoms, interstellar extinction, and temperature of the B~68~SMM cloud core -- an example of the evolution of physical conditions during the contraction Phase~1. After the contraction phase, these parameters were allowed to remain unchanged for a total integration time of 3.5Myr.}
 \label{att-phys}
\end{figure*}
We investigate ice chemistry of four starless cores with their central density $n_\mathrm{H}=n(\mathrm{H})+2n(\mathrm{H_2})$, cm$^{-3}$, and interstellar extinction $A_V$, mag, derived from observations. Only the core center part was considered (i.e., 0D model), where $n_{\rm H}$ and $A_V$ are at their highest. Each simulation consists of a collapse Phase~1, when the cloud density increases, and a quiescent Phase~2 with constant physical conditions.

A single-point 0D model was chosen because the ices in the center of each core most likely do exist, despite a number of uncertain parameters. These include the efficiency of non-thermal desorption mechanisms and turbulent motions in starless cores, as observed by \citet{Redman06,Levshakov14}, and \citet{Steinacker14}. Turbulence was not considered in the present study. Also, ices in core center likely have experienced the full range of physical conditions of the core since its formation. The simple 0D model puts certain constraints for the interpretation of results.

Cores with a Plummer-like density profile \citep{Plummer11,Whitworth01}, embedded in a surrounding cloud with a fixed column density $N_{\rm out}$, were modeled. The total column density to the core center can be calculated with
   \begin{equation}
   \label{phys-1}
N_\mathrm{H}=\int^0_{r_1}\frac{n_{\mathrm{H},0}}{[1+(r/r_0)^2]^{\eta/2}}\mathrm{d}r+N_\mathrm{out}, \mathrm{cm^{-2}},
   \end{equation}
where $n_{\rm H,0}$ is the (maximum) density at the center of the core, $r_0$ is the radius of the central density plateau, $\eta$ is the power-law slope at large radii, and $r_1$ is the radius of the core. The chemical modeling was done only for the center of the core, where $r=0$ and the density profile of the core is necessary only for the calculation of $N_H$ at its center. Table~\ref{tab-phys} summarizes the parameters $n_{\rm H,0}$, $r_0$, $r_1$, and $N_{\rm out}$. These are unique for each core, and were derived from \citet{Lippok13} for the CB~17, CB~26, CB~27, and B~68 cores and from \citet{Schmalzl14} for the CB~17 core. The two `versions' of CB~17 are hereafter denoted CB~17~S and CB~17~L, for data from \citeauthor{Schmalzl14} and \citeauthor{Lippok13}, respectively. This means that there are a total five models (CB~17~S, CB~17~L, CB~26, CB~27, and B~68) of four cores considered.

Table~\ref{tab-phys} data are relevant to cores in the stable Phase~2, when the contraction in Phase~1 has ended and the physical parameters of the cores are unchanging. This phase has an assumed integration time of $t_2>$2Myr for each modeled core. Prior to the stable phase, a core formation phase was considered. The formation of starless cores have been described before as a free-fall \citep[e.g.,][]{Lee04} or `delayed free-fall' \citep[e.g.,][]{Taquet14}. A `static', step-like contraction was employed by these authors.

In the present study, the formation of the core was described as a delayed free-fall process. The approach presented by \citet{Brown88}, \citet{Nejad90} and \citet{Rawlings92} was used. A Plummer-like core with an initial peak density of $n_{\rm H,ini}=3000$cm$^{-3}$ contracts to the final density $n_{\rm H,0}$ in a core formation time $t_1$ (i.e, length of Phase~1, see Table~\ref{tab-phys}). The parameter $B$ in equation~(1) of \citet{Rawlings92}, which modifies the rate of the contraction, was taken to be 2/3. This means that the collapse to the final densities, indicated in Table~\ref{tab-phys}, occurs in 1.4-1.5Myrs. Figure~\ref{att-phys} shows an example of the evolution of physical conditions for a core (B~68) during the contraction phase, up to 1.5Myr.

A value $B=0.7$ has been used by \citet{Nejad90}, \citet{Rawlings92}, and \citet{Ruffle99} for prestellar core models. We take $B=2/3$ as a somewhat lower value for starless cores. The exact value of $B$ has only a limited effect on ice composition, the main subject of the present study. This is because ice accumulation was calibrated by modifying the poorly-known efficiency of indirect reactive desorption to match the observed proportions of H$_2$O, CO, and CO$_2$ (section~\ref{ird}).

For the core formation epoch, $r_0$ was kept proportional to $n_{\rm H,0}^{-1/2}$ \citep{Keto10,Taquet14}. The Plummer-like density profile of the contracting core was retained, while the density increases from $n_{\rm H,ini}$ to $n_{\rm H,0}$. The value of $r_1$, which determines the physical size of the region in consideration, was assumed constant. This means that the mass of the core (i.e., the number of H atoms within a sphere of radius $r_1$) increases during the formation phase. $N_{\rm out}$ was retained constant, too.

The the total integration time $t_1+t_2$ was taken to be 3.5Myr and the results were analyzed for an interval of 1.0-2.3Myr (see section~\ref{r-obs}). In context with ice formation, the full time consists of three periods: (1) cloud with $n_{\rm H,0}\approx3\times10^3$cm$^{-3}$ and no ice on the grains; (2) ice accumulation up to a maximum thickness in a rapidly contracting core and up to 200kyr into the quiescent phase; (3) slow processing of ice by cosmic-ray induced photons after the freeze-out has ceased. The start of period (2) is marked with the appearance of first subsurface ice species at densities $5\times10^3$ to $1\times10^4$cm$^{-3}$. During the last period (3) the gas and the surface are in an approximate adsorption-desorption equilibrium. The nominal ice thickness decreases during period (3).

Ice photoprocessing leads to a declining amount of CO in the icy mantles. Because of this, quiescent Phase~2 integration times longer than 2Myr were not considered. For longer times, the H$_2$O:CO:CO$_2$ ice abundance ratio is no longer supported by observations of interstellar ices \citep{Gibb04,Whittet07,Boogert11,Oberg11b}. Mantle growth and the abundances of major ice species are considered in section~\ref{majres}.

Gas and dust were assumed to be in thermal equilibrium with temperature derived from the interstellar extinction $A_V$, in line with \citet{Garrod11}. In other words, it was assumed that the cores are heated externally by the interstellar radiation field. $A_V$ was calculated with:
   \begin{equation}
   \label{phys0}
A_V=\frac{N_\mathrm{H}}{1.60\times10^{21}},
   \end{equation}
Table~\ref{tab-phys} shows the calculated values for $A_V$ and $T$ for each of the modeled cores in the stable phase.

Table~\ref{tab-ab} shows the chemical abundances used at the start for all simulations. They are based on the abundances derived by \citet{Garrod06} and \citet{Wakelam08}. The rate of hydrogen ionization by cosmic rays was taken to be $1.3\times10^{-17}$s$^{-1}$. The flux of cosmic-ray induced photons and interstellar photons was taken to be 4875 and $10^8$cm$^{-2}$s$^{-1}$, respectively, with a grain albedo of 0.5.

The self- and mutual-shielding of H$_2$, CO, and N$_2$ was included with the use of the tabulated shielding functions \citep{Lee96,Li13}. Shielding was also attributed to molecules in ice. It is important to note that the surrounding molecular cloud was included in these calculations by assuming an external H$_2$ column density of $N_{\rm out}/2$. This was not included in Papers I and II that used a physical model built on similar principles. Such more accurate approach here is of particular importance, because more plausible values for indirect reactive desorption efficiency can now be derived (section~\ref{ird}). The synthesis of hydrogen on grain surface was considered by, both, Langmiur-Hinshelwood and Eley-Rideal mechanisms, although the latter is of little importance.

\subsubsection{Dust grain properties}
\label{grdust}

The grains were assumed to consist of two components. The first is inert, solid nuclei with a radius of $a=0.1\mu$m and an abundance $1.32\times10^{-12}$ relative to H atoms. In dark and dense conditions ($A_V>1.9$mag), the second component, interstellar ice, begins to accumulate onto grain surfaces in significant amounts. The increase of grain size is calculated self-consistently, assuming a thickness of 350pm for each ML, which is multiplied by the number of ice MLs to obtain ice thickness $b$. The number of adsorption sites per ML was retained constant at $1.5\times10^6$. Changes in this number likely do not influence the agreement of model results with observations \citep{Acharyya11,Taquet12}, and were not considered in the model. Dust grain albedo was assumed to be 0.5.

The ice layer on the grains was described with the sublayer approach, developed in \citetalias{Kalvans15b}. This means that a fully formed  ice (accumulation from gas has ceased) consists of a surface layer (1-2 MLs) and three subsurface layers -- `sublayers' -- of approximately equal thickness. A sublayer nominally may contain up to several tens of MLs. Two or more sublayers in a model allow to consider the depth-dependent composition of the ice mantles. Three sublayers were considered in the present study, similarly to \citetalias{Kalvans15b}. For reference, they were numbered with the first being the shallowest and the third being the sublayer adjacent to grain nucleus. The transition from surface to the first sublayer and, subsequently, to second and third sublayers has been harmonized with the accretion rate, in line with Papers I and II.

\subsubsection{Ice physical description}
\label{pice}

The molecules are bound to the surface or into the mantle with their adsorption ($E_D$) or absorption energies ($E_B$), respectively. $E_D$ is the `basic value', given in the reaction network, from which all the other energies were calculated. The binding energies, used for calculating the diffusion rates, for surface ($E_{b,S}$) and mantle ($E_{b,M}$) species were assumed to be 0.40$E_D$ and 0.40$E_B$, respectively \citep{Garrod11}. In Papers I and II it was argued that $E_{b,M}$ has to be higher, or similar to $E_D$, otherwise the movement of molecules in ice bulk would be faster than their evaporation from surface, which fits the description of a liquid, not solid. In line with the earlier works, it was assumed that $E_B=3E_D$, i.e., that typically $E_{b,M}=1.2E_D$. For CO and CO$_2$ molecules, the ratios $E_{b,S}/E_D=E_{b,M}/E_B$ were taken to be 0.31 and 0.39, respectively \citep{Karssemeijer14}.

In dark cores, the grains may be sufficiently cold to have a significant proportion of their surface covered by molecular hydrogen. H$_2$ provides a much weaker binding than water ice. The adsorption energy for H and H$_2$ was calculated according to \citet{Garrod11}. For other species, $E_D$ was not affected by the surface coverage of H$_2$ because of the large difference between the hopping rate of H$_2$ and vibration frequency of heavy molecules \citepalias{Kalvans15b}.

Accretion of neutral gaseous species onto the grains with a radius $a+b$ was calculated with sticking probabilities of 0.33 for light species (H, H$_2$) and 1.0 for all other species \citep{Kalvans10}. A total of six desorption mechanisms were considered. The rate coefficients for evaporation and cosmic-ray-induced whole-grain heating \citep{Hasegawa93a} were calculated for each molecule as a thermal desorption at the equilibrium temperature and 70K, respectively. Desorption by interstellar and cosmic-ray-induced photons was considered with uniform yields for all species, 0.003 and 0.002, respectively. These values were chosen to be in agreement with recent photodesorption experiments with interstellar ice analogs, as discussed in \citetalias{Kalvans15b}. Photodesorption from subsurface mantle species was permitted, if the number of overlying MLs does not exceed two. The yield for subsurface photodesorption was reduced by a factor of $1/3$. This approach was developed to be in line with the results of \citet{Andersson08}.

A reaction-specific reactive desorption was used, with the $\alpha_{\rm RRK}$ parameter set to 0.03 \citepalias{Kalvans15b}. This is the ratio of the surface-molecule bond frequency to the frequency at which energy is lost to the grain surface \citep{Garrod07}.

\subsubsection{Indirect reactive desorption}
\label{ird}
\begin{table*}
\begin{center}
\footnotesize
\caption{Comparison of observed and calculated (B~68 core) CO and CO$_2$ ice abundances, relative to water. Observational data from \citet{Whittet07}, unless noted otherwise.}
\label{tab-obs}
  \begin{tabular}{llccccc}
\tableline
\tableline
 & & \multicolumn{2}{c}{Observations} & \multicolumn{2}{c}{B 68 model} & \\
Source ID & $A_V$ & $\rm \frac{CO}{H_2O}$\tablenotemark{a},\% & $\rm \frac{CO_2}{H_2O}$,\% & $\rm \frac{CO}{H_2O}$,\% & $\rm \frac{CO_2}{H_2O}$,\% & $t$\tablenotemark{b}, kyr \\
\tableline
043728.2+261024 & 6.3 $\pm$ 1.5 & . . . & 25.0 & 1.8 & 13.8 & 1417 \\
042324.6+250009 & 10.0 $\pm$ 0.5 & 20.2 & 19.1 & 7.7 & 13.0 & 1451 \\
043325.9+261534 & 11.7 $\pm$ 0.5 & 12.0 & 16.7 & 12.5 & 13.9 & 1459 \\
043926.9+255259 & 15.3 $\pm$ 0.5 & 27.3 & 16.7 & 30.2 & 16.5 & 1471 \\
042630.7+243637 & 17.8 $\pm$ 1.5 & 45.1 & 18.3 & 43.2 & 18.2 & 1476 \\
\tableline
\multicolumn{2}{c}{Threshold ice abundances} & & H$_2$O, ML & CO$_2$, ML & CO, ML & \\
$A_{\rm th}$ H$_2$O & 3.2 $\pm$ 0.1 & & 3.5 & & & 1332 \\
$A_{\rm th}$ CO$_2$ & 4.3 $\pm$ 1.0 & & & 1.5 & & 1375 \\
$A_{\rm th}$ CO\tablenotemark{c} & 6.8 $\pm$ 1.6 & & & & 0.6 & 1424 \\
$A_{\rm th}$ CO\tablenotemark{d} & 8.1 & & & & 1.3 & 1437 \\
\tableline
\end{tabular}
\tablenotetext{a}{Abundance ratio for icy species}
\tablenotetext{b}{Integration time, corresponding to the $A_V$ value}
\tablenotetext{c}{\citet{Whittet01}}
\tablenotetext{d}{\citetalias{Kalvans15b}}
\end{center}
\end{table*}
Desorption induced by the energy released from the H+H exothermic surface reaction probably determines the relative proportions of major ice components H$_2$O, CO, and CO$_2$. In order to reproduce observations of interstellar ices, a sequence $\rm CO > CO_2 >> H_2O$ must be observed for indirect reactive desorption efficiency $f_{\rm H_2fd}$ \citepalias{Kalvans15b}. Without indirect reactive desorption, the calculated abundance of solid carbon species in collapsing cores can be overestimated, when compared to the observational results provided by \citet{Gibb04,Whittet07}, and \citet{Oberg11b}. For example, realistic CO and CO$_2$ relative abundances can be achieved, if the ice contains a substantial amount of CH$_4$, H$_2$CO, and (or) CH$_3$OH \citep{Garrod11,Taquet14}. Of the latter three species only methanol has been occasionally observed in interstellar ices \citep{Boogert11}.

The parameter $f_{\rm H_2fd}$ can be described as the number of desorbed molecules per H$_2$ formed on a grain. Exact value for the efficiency are unknown, and, given the averaging and uncertainties in astrochemical numerical simulations, it is reasonable to adjust $f_{\rm H_2fd}$ so that it produces an approximate fit with observations of starless cores. Such an approach can be used until a more precise value of $f_{\rm H_2fd}$ is found in experiments or, for example, quantum chemistry calculations.

In order to derive $f_{\rm H_2fd}$ for the present study, we employ a similar method to \citetalias{Kalvans15b}. A prestellar core in fee-fall gravitational collapse was considered in that study. For the present research, the B~68 core model (Table~\ref{tab-phys}) was used. It has the highest final density and final $A_V$ among all the starless cores in consideration. This allows the comparison of observed/modeled CO:H$_2$O and CO$_2$:H$_2$O ice abundance ratios for five $A_V$ values from the dataset of \citet{Whittet07}. An additional observational constraint for the model is the abundance of H$_2$O, CO, and CO$_2$ ices at their respective threshold extinctions $A_{\rm th}$.

The above means that, while the B~68 cloud core is currently stable \citep{Bergin06,Redman06}, we assume that it formed in a delayed free-fall process, which ceased when the core reached its present density. This assumption has been attributed to all simulated cores in this study. While it might not be entirely correct, it is probably just as useful as other realistic assumptions, because of the poorly constrained value of $f_{\rm H_2fd}$, which has to be calibrated, regardless of the exact core formation path.

Table~\ref{tab-obs} summarizes the observational and best-fit B~68 model results. The desorption efficiency $f_{\rm H_2fd}$ was calculated according to the empiric approach of \citetalias{Kalvans15b}:
   \begin{equation}
   \label{phys4}
f_\mathrm{H_2fd}= Q \mathrm{exp}(-E_D/(Q_1),
   \end{equation}
where $Q=9.5\times10^{-5}$ and $Q_1=10^3$. The adopted best-fit values of $Q$ and $O_1$ are different from those in \citetalias{Kalvans15b}. It is possible that the values obtained here are more reliable than those of \citetalias{Kalvans15b} because of the more advanced and realistic physical model (section~\ref{ccond}).

The observational values for CO:H$_2$O and CO$_2$:H$_2$O ice abundance ratios were taken from \citet{Whittet07}. While newer datasets are available \citep{Boogert11,Oberg11b}, \citeauthor{Whittet07} provides the $A_V$ for each background star, which enables to tie the observations to specific time-points of the simulations with the corresponding $A_V$ \citepalias[see][for more on this approach]{Kalvans15b}.

\subsection{Chemical model}
\label{chem}

\subsubsection{Reaction network}
\label{reacs}
\begin{table*}
\begin{center}
\caption{Changes in reactions respective to the original network by \citet{Laas11}.}
\label{tab-reac}
\begin{tabular}{clcccl}
\tableline\tableline
No. & New gas phase reactions\tablenotemark{a} & $\alpha$ & $\beta$ & $\gamma$ & Reference \\
\tableline
1 & $\rm CH_2OH + H \longrightarrow CH_3 + OH$ & 1.6E-10 & 0 & 0 & NIST \\
2 & $\rm CH_2OH + O \longrightarrow H_2CO + OH$ & 1.5E-10 & 0 & 0 & NIST \\
3 & $\rm CH_2OH + CH_3 \longrightarrow H_2CO + CH_4$ & 1.4E-10 & 0 & 0 & NIST \\
4 & $\rm CH_2OH + CH_3 \longrightarrow C_2H_5OH$ & 1.0E-15 & -3.00 & 0 & \citet{Vasyunin13b}\tablenotemark{b} \\
\tableline
No. & New surface (ice) reactions & $E_A$, K &  &  & Reference \\
\tableline
5 & $\rm H + HCO \longrightarrow CO + H_2$ & 0 &  &  & This work \\
6 & $\rm OH + HCO \longrightarrow CO + H_2O$ & 0 &  &  & This work \\
7 & $\rm NH_2 + HCO \longrightarrow CO + NH_3$ & 0 &  &  & This work \\
8 & $\rm H_2 + HCO \longrightarrow CH_3O$ & 2100\tablenotemark{c} &  &  & This work \\
9 & $\rm OH + SO \longrightarrow SO_2 + H$ & 0 &  &  & Gas phase \\
10 & $\rm O + HS \longrightarrow S + OH$ & 956 &  &  & Gas phase \\
11 & $\rm O + HS \longrightarrow SO + H$ & 0 &  &  & Gas phase \\
12 & $\rm H + CL \longrightarrow HCL$ & 0 &  &  & Gas phase \\
\tableline
No. & Changed surface (ice) reactions & $E_A$, K &  &  & Reference \\
\tableline
13 & $\rm CO + O \longrightarrow CO_2$ & 290 &  &  & \citet{Roser01} \\
14 & $\rm O_2 + H \longrightarrow O_2H$ & 600 &  &  & \citet{Du12} \\
15 & $\rm H + H_2CO \longrightarrow CH_3O$ & 2100 &  &  & \citet{Woon02} \\
16 & $\rm O_3 + H \longrightarrow O_2+OH$ & 0 &  &  & \citet{Mokrane09} \\
\tableline
\end{tabular}
\tablenotetext{a}{In addition to the four reactions from \citet{Vasyunin13b}. Reaction rate coefficient $k = \alpha (T/300)^\beta \mathrm{exp}(-\gamma/T)$}
\tablenotetext{b}{Data adopted from $\rm CH_3O + CH_3$ reaction}
\tablenotetext{c}{Estimate}
\end{center}
\end{table*}
Several changes have been introduced into the gas-grain network, which was acquired from \citet*{Laas11}. The four new gas-phase reactions of \citet{Vasyunin13b} were added. All these reactions involve the methoxy radical CH$_3$O. Because these authors did not discern between the methoxy radical and its isomer, the hydroxymethyl radical CH$_2$OH, four complementary reactions were added for the latter species (1-4 in Table~\ref{tab-reac}).

The results of \citetalias{Kalvans15b} show that formic acid HCOOH, formaldehyde H$_2$CO, and formamide NH$_2$CHO can be overproduced in subsurface mantle conditions with the surface reaction set of \citet{Laas11}, which has been adopted from \citet{Garrod08}. In an attempt to reduce their abundance, grain surface reactions 5-8 in Table~\ref{tab-reac} have been added to the network. They tend to reduce the production efficiency of HCOOH, H$_2$CO, and NH$_2$CHO because alternative outcome is now possible for their reactants.

Reactions 5-7 were assumed to be barrierless. Reaction~8 has an assumed activation barrier $E_A$=2100K, estimated from reactions of H$_2$ with compounds that have a similar structure to HCO. The barrier of the latter reaction is so high that it has little effect on the results of the current cold core model.

Reactions 9-11 were added to more adequately reproduce the chemistry of sulfur oxides, with activation barriers taken from similar gas-phase reactions. Finally, reaction 12 was added to avoid most of chlorine atoms ending up in solid ClO, which is probably unrealistic, given the availability of atomic hydrogen, irradiation, and an aqueous medium on grain surfaces. The activation barriers for four additional surface reactions -- 13-16 in Table~\ref{tab-reac} -- were also changed, in line with available data.

\subsubsection{Surface and bulk ice chemistry}

The rate for reactions on grain surface was calculated according to the reaction-diffusion competition approach of \citet{Garrod11}. The modified rate equations method has been applied \citepalias[see][]{Kalvans15b}. As explained in Section~\ref{pice}, part of surface reaction products desorb into the gas. Reaction barriers are overcome either thermally or by quantum tunneling, whichever is faster \citep{Garrod08}.

The rate of binary reactions in bulk ice was calculated for reactants that are immobile for most of the time (Papers I and II). It was assumed that each molecule vibrates in a lattice cell with ten neighboring molecules. Reactants have to overcome a certain energy barrier $E_{\rm prox}=0.1E_D$ in order to achieve sufficient proximity for a reaction to occur. Molecule diffusion in ice and reaction-diffusion competition have been taken into account in a manner, similar to that for surface reactions. I refer the reader to Papers I and II for a more detailed justification and formalism for the calculation of reaction rate coefficients.

Surface and bulk-ice molecules are subjected to photodissociation by interstellar and cosmic-ray-induced photons. For surface species, the dissociation rate has been taken to be equal to that of gas-phase species. For molecules within the mantle, the photon flux is attenuated by overlying MLs. Each ML has an assumed absorption probability of 0.007 \citep{Andersson08}. In order to calculate the photon flux, respective to each sublayer, the value relevant to middle ML of that sublayer was used \citepalias[see][]{Kalvans15b}.

\section{Results: ice formation and major components}
\label{majres}

To avoid repetition in the studies of five roughly similar models of objects of the same class, the modeling results are not reviewed for each case separately. Instead, we focus on several astrochemical problems, relevant for ices in starless cores. The use of five models with physical conditions derived from observations instead of an abstract model adds to the credibility of the research. We also have to keep in mind that the 0D model adequately represents only the center part of each core. The abundances values used here are calculated with N(\textit{X})/N(H), i.e., relative to total hydrogen. This means that observationally detected abundances that are given relative to H$_2$ were divided by two for comparison with calculation results.

\subsection{The evolution of the ice layer}
\label{r-mform}
\begin{table*}
\begin{center}
\caption{Calculated ice composition parameters in the starless core models.}
\label{tab-thick}
\begin{tabular}{rr|c|cccc|c|ccc}
\tableline\tableline
 &  & $t_{\rm sub}$, & \multicolumn{4}{c}{Max. ice thickness} & $t_{50}$, & \multicolumn{3}{c}{Ice at 3.5Myr} \\
Core & $A_V$ & Myr & ML & $t_{\rm max}$, Myr & $\rm \frac{CO}{H_2O}$, \% & $\rm \frac{CO_2}{H_2O}$, \% & Myr & ML & $\rm \frac{CO}{H_2O}$, \% & $\rm \frac{CO_2}{H_2O}$, \% \\
\tableline
CB 17 S & 13.8 & 1.25 & 170 & 1.57 & 81 & 23 & 2.31 & 147 & 68 & 81 \\
CB 17 L & 9.1 & 1.21 & 164 & 1.62 & 74 & 35 & 1.92 & 141 & 61 & 107 \\
CB 26 & 6.5 & 1.21 & 153 & 1.68 & 69 & 59 & 1.57 & 133 & 48 & 150 \\
CB 27 & 9.5 & 1.06 & 158 & 1.60 & 69 & 51 & 1.58 & 131 & 51 & 175 \\
B 68 & 17.6 & 1.15 & 171 & 1.52 & 78 & 25 & 2.12 & 146 & 63 & 90 \\
\tableline
\end{tabular}
\tablecomments{$t_{\rm sub}$--time for ice thickness to reach 1ML; $t_{\rm max}$--time of maximum ice thickness; $t_{50}$--time for the CO$_2$:H$_2$O ice abundance ratio to reach 50\%.}
\end{center}
\end{table*}
\begin{figure*}
 \vspace{-2cm}
  \hspace{-1cm}
  \includegraphics[width=18.0cm]{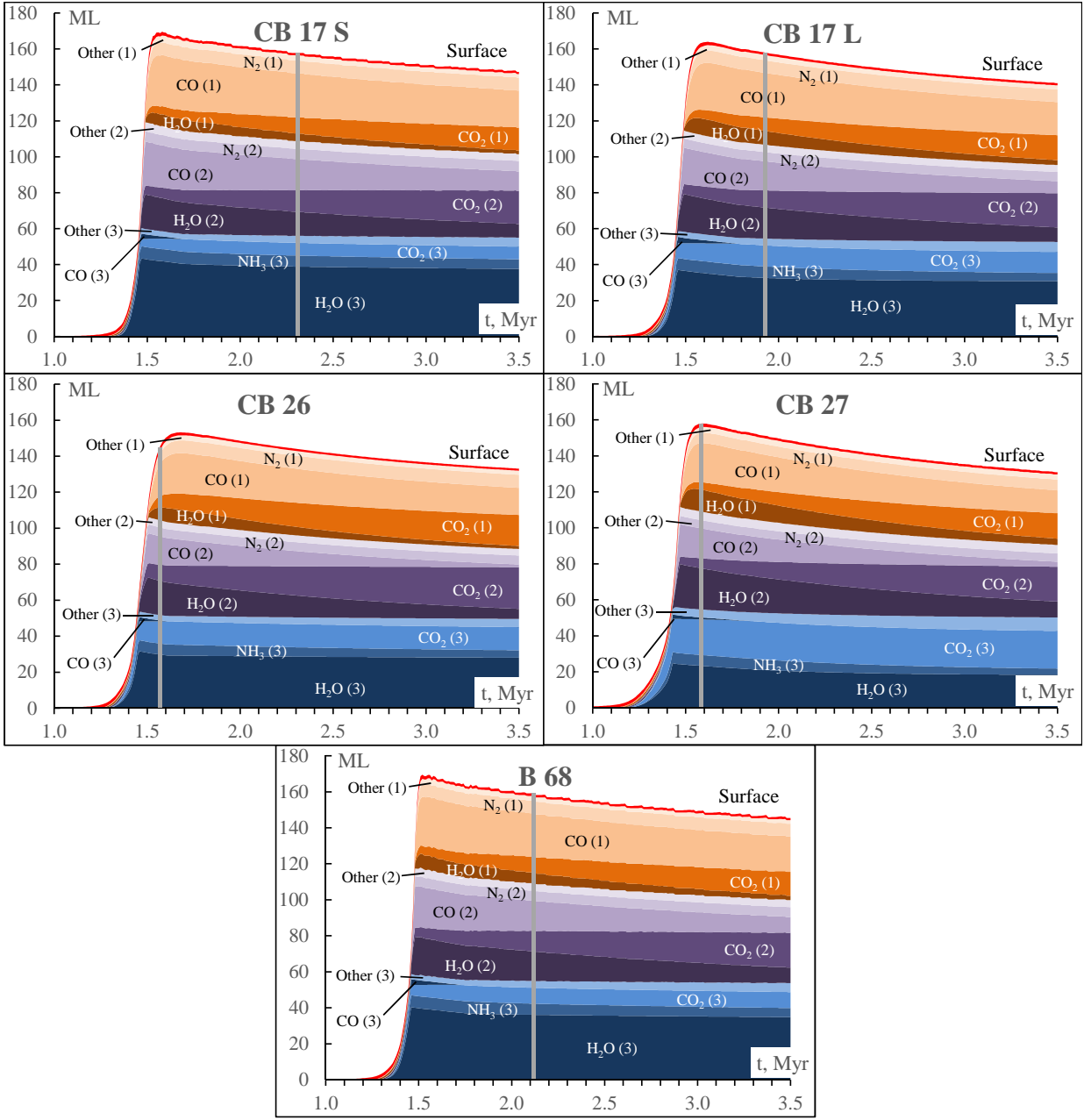}
 \vspace{-7cm}
 \caption{Calculated ice thickness and abundance in MLs for major species in the sublayers for the modeled cloud cores. The number in parenthesis indicates the number of the respective sublayer, while `Other' stands for all other icy species in that sublayer. The vertical gray line indicates $t_{50}$ for each model.}
 \label{att-sub}
\end{figure*}
\begin{figure*}
  \includegraphics[width=18.0cm]{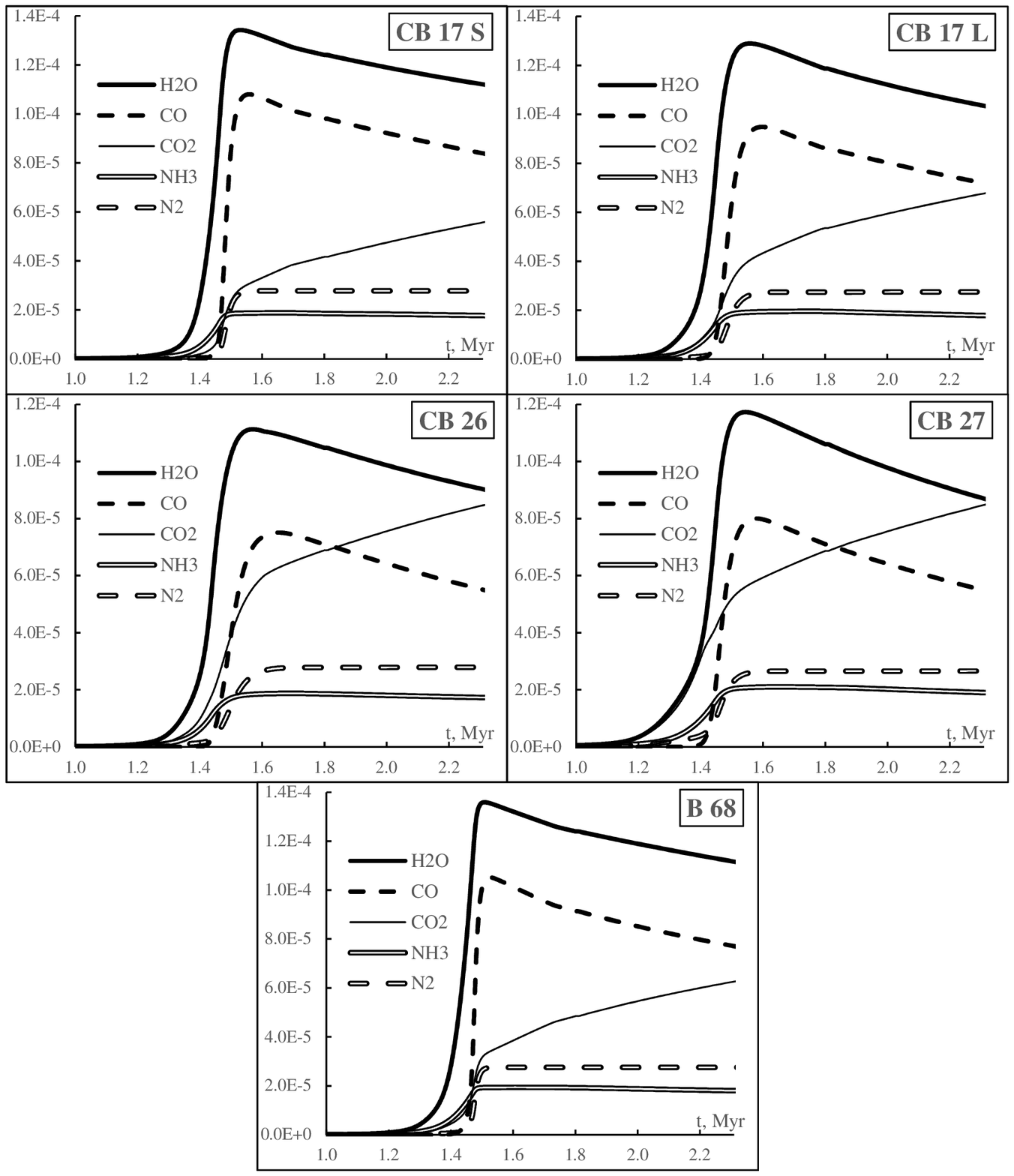}
 \vspace{-8cm}
 \caption{Calculated abundances, relative to H atom number density, of major species in ice for the five modeled cores.}
 \label{att-gen}
\end{figure*}
Ice layer is basically nonexistent (around 0.1ML) when $A_V<1.9$mag, which means that the calculation results (abundances of species) are irrelevant for ice chemistry if $t\le1$Myr. Because of this, we interpret results for integration times $t>1$Myr, only. Ice formation observes the rules outlined in \citetalias{Kalvans15b}. As the cloud contracts and $A_V$ increases over 2mag, ice molecules begin to accumulate on grain surfaces, forming a mixture of H$_2$O, CO$_2$, and NH$_3$. The ice is dominated by interstellar photons, which means that all surface CO is oxidized to CO$_2$ via the sequence:
   \begin{equation}
   \label{res1}
\mathrm{H_2O} (h\nu) \rm O,OH (CO) CO_2 ,
   \end{equation}
probably the most important chemical transformation in bulk ice. When the cloud becomes darker, the flux of interstellar photons is diminished, rapid ice photoprocessing becomes impossible, and CO begins to accumulate at $A_V>5$. The accretion of CO is faster than the surface synthesis of CO$_2$, and CO becomes the second most abundant ice molecule when $A_V>12$mag.

Figure~\ref{att-sub} shows the nominal thickness of ice and the abundances in MLs for a few most important species in each sublayer for all five starless cores. Ice thickness peaks shortly after the end of the core collapse stage, and then slowly decreases. The reason for this decrease is the conversion of simple ice species into more complex ones via photoprocessing. Most importantly, the sequence~(\ref{res1}) converts water and carbon monoxide into carbon dioxide. Thus, the number of molecules per grain and the nominal ice thickness are reduced. This can be seen in figure~\ref{att-gen}. A significant amount of hydrogen is released in the process. H$_2$ partially accumulates in the mantle when $T<10K$. The inner sublayer~3 is the least affected because it has a lower proportion of CO and it is better shielded than the above layers. These results are different from those of \citet{Chang14} thanks to the gradual increase of gas density and $A_V$ during Phase~1, which ensure an ice composition that is much more differentiated between layers. Such a chemical differentiation (H$_2$O in inner and CO in outer layers) has been confirmed by models that consider ice formation in collapsing cores with monolayer accuracy \citep{Garrod11,Taquet14}.

CB~26 is the smallest of the five cores, it has the lowest final $A_V$ and density values (Table~\ref{tab-phys}). This means that molecules on grains deposit at a lower rate, and have undergone significant subsurface photoprocessing even before the freeze-out has ended.  With the `freeze-out epoch' we understand the interval between the time of formation for the first sublayer ($t_{\rm sub}$) and the time, when maximum ice thickness is reached ($t_{\rm max}$, as specified in table~\ref{tab-thick}). $t_{50}$ is the point in time when the CO$_2$:H$_2$O ratio in ice has grown to 50\%, and for CB~26 it occurs even before $t_{\rm max}$.

On the other hand, CB~27 is the largest core considered. Paradoxically, this results in a similar ice composition to that of CB~26. The $A_V$ threshold for accumulation of a significant mass of ice on the grains ($>$1ML) is close to 1.9mag. This occurs at a density of $\approx5\times10^3$cm$^{-3}$ for CB~27, and in the range $(8-9)\times10^3$cm$^{-3}$ for the other four cores. The result is that the first ice sublayers in CB~27 appear some 150kyr earlier than in other cores. The formation of sublayer~1 in CB~27 is complete some 20kyr earlier than for the other cores. This means that sublayer~1 has formed in a relatively diffuse environment, which is still largely dominated by interstellar photons, where water photodissociation products combine with CO on the surface to produce an ice that contains a large proportion of CO$_2$.

Because indirect reactive desorption, which largely governs ice composition, has been calibrated for the B~68 core (section~\ref{ird}), the explained mechanisms may lead to different results, if $f_{\rm H_2fd}$ is different. The important thing is that both, a diffuse medium or early ice formation, may lead to a significantly photoprocessed ice. The first is characteristic for low-mass molecular cloud cores, while the second is more likely for relatively massive cores.

Table~\ref{tab-thick} summarizes data for each model at two important points in time: when ice thickness is at its maximum, and at the end of the simulation run. The thickest ice layer is attained in CB~17~S and B~68 models. High ice thickness for B~68 can be expected because of its high extinction and final density. For CB~17~S, neither its density, nor extinction are the highest. A thick ice layer for this core arises because $N_{\rm out}$ for this model is 0, which means that interstellar photons are attenuated to a lesser extent and the ice layer begins to form at a later, denser stage of the collapsing core. In turn, ice accumulates rapidly and undergoes less intense photoprocessing, and a smaller number of CO and H$_2$O molecules are converted to CO$_2$. This results in higher nominal ice thickness. This conclusion is supported by the abundance of carbon oxides -- a higher proportion of CO$_2$ means a nominally thinner ice (Table~\ref{tab-thick}).

The above analysis confirms that extinction largely regulates the formation of the ice layer on the grains because $A_V$ determines the number of interstellar photons that reach the grains \citep{Watson72}. Interstellar extinction also affects the abundance of radical species on surfaces and in the gas, which determines the efficiency of direct and indirect reactive desorption. It can be safely said that the formation history of a starless or prestellar core determines its ice composition.

\subsection{General ice chemistry}
\label{r-gen}

Aside water and carbon oxides, the nitrogen species NH$_3$ and N$_2$ also can be considered major ice constituents. Their abundance is little affected by ice photoprocessing. Other ice species whose abundance may exceed 1\% relative to water for at least one of the modeled cores include H$_2$, O$_2$, H$_2$O$_2$, O$_2$H, and H$_2$S.

Figure~\ref{att-sub} shows ice evolution up to 3.5Myr, the full length of the simulation runs. Although the modeled cores have different physical parameters, ice sublayers share a similar composition in all cases. The inner sublayer~3 contains a large amount of water, in addition to CO$_2$ and NH$_3$. All CO molecules here are quickly consumed, which makes chemical processes in sublayer~3 somewhat different to those in above layers. The lack of CO and its daughter atomic C means that O, OH, N, NH, S, and other radicals are more available.

Sublayer~2 initially forms as a $\approx$1:1 mixture of H$_2$O and CO, with additions of CO$_2$ and N$_2$. As the time progresses, CO and H$_2$O combine, and the sublayer becomes increasingly dominated by CO$_2$. The outer sublayer~1 largely consists of CO, with an admixture of H$_2$O, CO$_2$, and N$_2$ at roughly similar proportions. CO remains the main species in this layer, although the proportion of CO$_2$ is steadily growing. Sublayer composition may profoundly affect the chemistry of minor species because the main reacting species are photodissociation fragments of major molecules.

\subsection{Comparison with observations and age limits}
\label{r-obs}
%
\begin{figure*}
  \includegraphics[width=18.0cm]{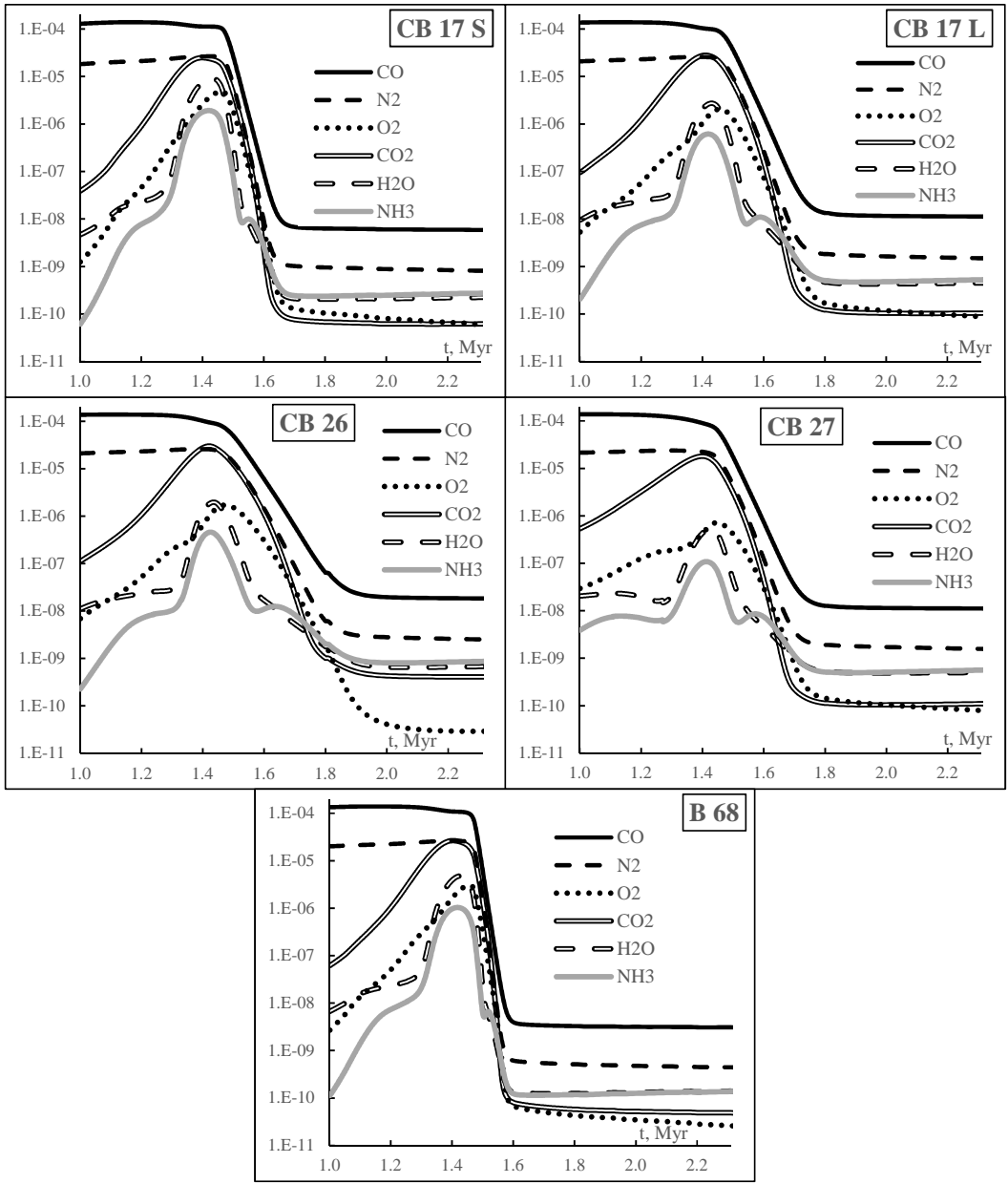}
 \vspace{-7cm}
 \caption{Calculated abundance of major gas phase molecules, relative to hydrogen.}
 \label{att-gas}
\end{figure*}
The initial chemical composition of ices is largely regulated by indirect reactive desorption, calibrated with the B~68 model (section~\ref{ird}). Table~\ref{tab-obs} shows that carbon oxides are somewhat depleted during the earlier stages in cloud contraction, compared to the observational data by \citet{Whittet07}. We found it impossible to remove this discrepancy with adjustments in the efficiency of indirect reactive desorption, Equation~(\ref{phys4}). This depletion may arise, for example, if the calculated abundance of atomic H is higher by a factor of two than H abundance in the real core. Another possible cause can be changes in surface properties as proportionally more and more CO molecules stick to the grains during the ice formation epoch. We will investigate the latter suggestion in a separate paper.

Figure~\ref{att-gas} shows that molecules, typically formed on grain surfaces or dense gas (e.g., H$_2$O, CO$_2$, NH$_3$, and O$_2$), are also abundant in the gas phase during or immediately after the core contraction stage, when density is in excess of $\approx10^4$cm$^{-3}$. Gas and surface molecules are actively interchanged during the freeze-out epoch. Such gas-phase abundance peaks in a layer around a collapsing core have been found in models of prestellar cores \citep[][; \citetalias{Kalvans15b}]{Garrod06,Vasyunin13a} and observations of water vapor in the prestellar core L1544 \citep{Caselli12}. The present model predicts is that such peaks are likely to be shorter and higher thanks to more efficient desorption mechanisms. Desorption retains a substantial amount of refractory species in the gas phase until very late times in core contraction.

We can conclude that the Phase~1 modeling results confirm that gas falling into a prestellar or protostellar core is rich with molecules synthesized on the surface. The abundance peaks predicted by the present study are higher than in other similar models, thanks to more efficient desorption mechanisms. The main desorption agent are interstellar photons, as suggested by \citet{Dominik05}.

Subsurface ice processing may place constraints for the maximum age of starless cores. The observed solid CO$_2$:H$_2$O abundance ratio does not exceed 44\% \citep{Boogert11}, although higher values are permitted by upper limits. As a threshold value for the age of the cores we chose the longest time taken for any of the cores to reach a CO$_2$:H$_2$O ice abundance ratio of 50\%, corresponding to a time $t_{50}$. For the CB~17~S core $t_{50}=2.31$Myr, longer than $t_{50}$ for other cores (Table~\ref{tab-thick}). A single end time of 2.31Myr was applied for all simulations. The decision to adopt a single and rather long total integration time for all models results in that for the CB~26 and CB~27 cores CO$_2$:H$_2$O is higher than 50\% for most of the time. With this approach the model considers a variety of icy environments and allows for an easy comparison between the five simulation results.

The above means that the actual lifetime $t_2$ of a quiescent Phase~2 core with constant physical conditions is below 0.9Myr for all models. This is significantly less than the initial assumption of $t_2\approx$2Myr for a 3.5Myr total simulation length. Thus, the modeling results were analyzed for an integration time interval of 1.00-2.31Myr. 

If gas and dust in the center part of a starless core is not effectively mixed with its surroundings, CO$_2$-rich ices should be present in starless cores. Turbulence and mixing of gas in the dense cores is a poorly-known parameter that may contribute to the destruction of the ices \citep{Boland82,Williams84,Redman06,Adams07,Levshakov14,Steinacker14}. Given the uncertainties, we treat the chemistry of CO$_2$-rich ices as a theoretical possibility. Their observational detection or non-detection might provide additional constraints on the physico-chemical processes that govern ice formation, chemical processing, and desorption.

\section{Results: elemental chemistry}
\label{elres}

\subsection{Oxygen chemistry in ice}
\label{r-o}

\subsubsection{Chemical processing of minor oxygen species}
\begin{figure*}
  \includegraphics[width=18.0cm]{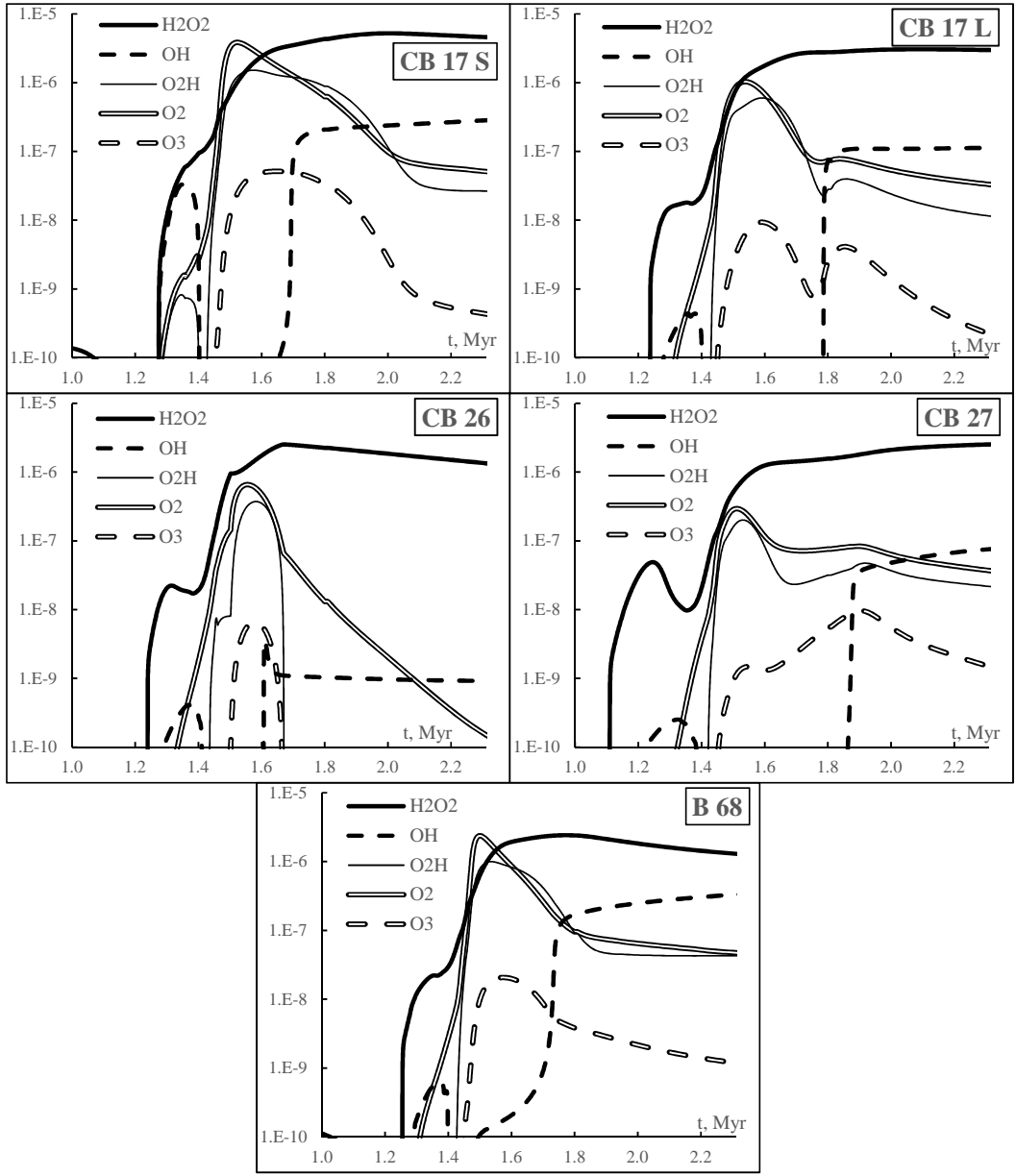}
 \vspace{-8cm}
 \caption{Calculated abundance of selected oxygen species in ice, relative to hydrogen.}
 \label{att-o}
\end{figure*}
\begin{figure*}
  \includegraphics[width=18.0cm]{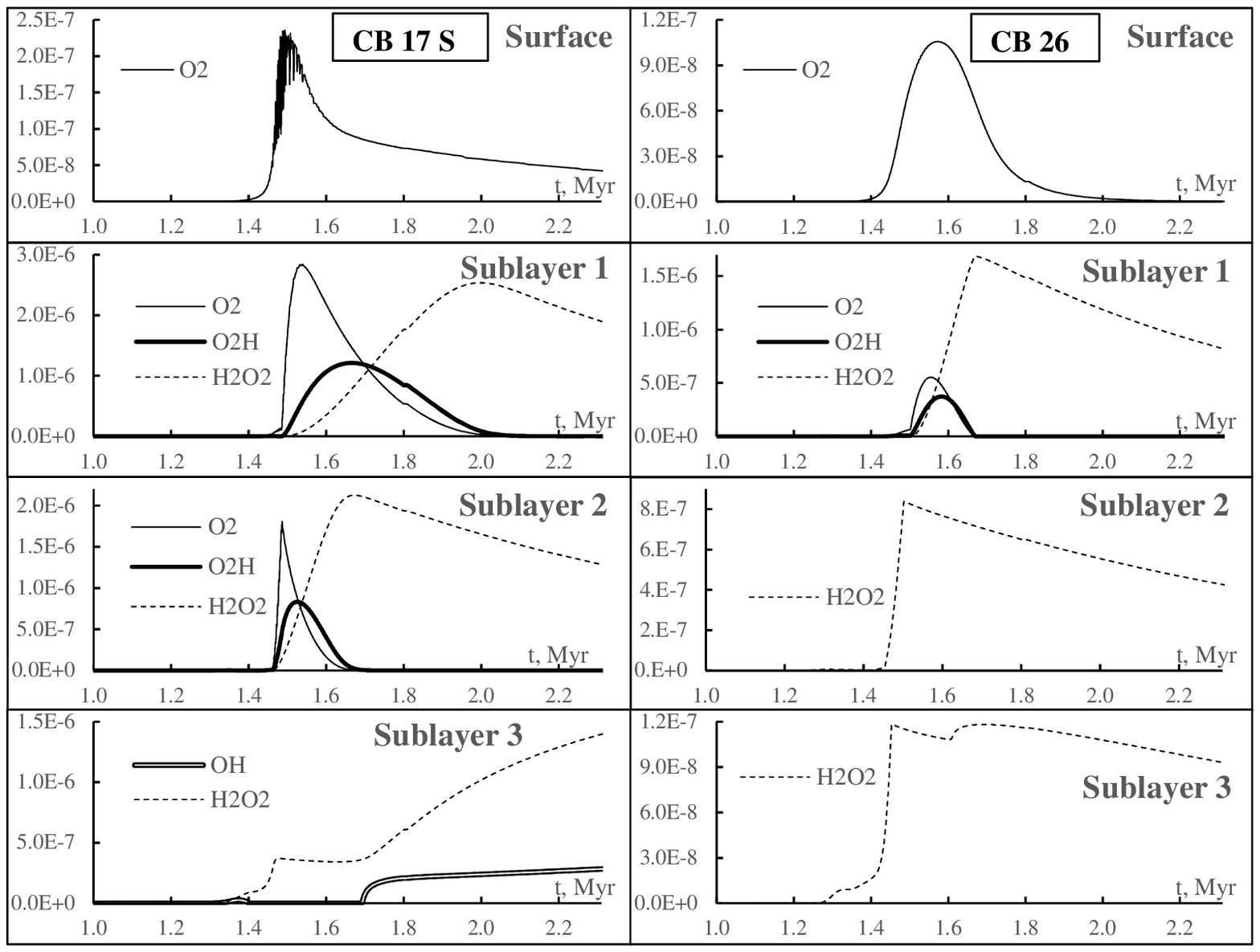}
 \vspace{-13.5cm}
 \caption{Calculated relative abundance of main O and O-H species in ice mantle sublayers for CB~17~S, and CB~26 models. The latter represents a significantly photoprocessed ice. The irregularities for CB~17~S surface layer at $t\approx1.5$Myr are artifacts.}
 \label{att-o-sub}
\end{figure*}
Three major ice species -- H$_2$O, CO, and CO$_2$ -- contain copious amounts of the eighth element, and the chemistry of oxygen in ice is closely connected to these molecules. When water and carbon oxides are excluded, the bulk ice chemistry of oxygen for CB~17~S and B~68 is initially dominated by molecular oxygen O$_2$, which is rather quickly converted into hydrogen peroxide H$_2$O$_2$, as shown in figure~\ref{att-o}. For other models, where ice photoprocessing is more intense, H$_2$O$_2$ is the more important species from the beginning. H$_2$O$_2$ is more stable in irradiated interstellar ices than O$_2$, which mainly arrives from the gas phase. The hydroperoxyl radical O$_2$H is the most abundant radical for oxygen chemistry in ice, thanks to its high $E_b,M$. It is generated by two mechanisms -- H addition to O$_2$ (early times) and the photodissociation of H$_2$O$_2$ (late times). Hydroxyl radical OH begins to accumulate to a limited extent in the inner sublayer~3 only after the CO molecule has been oxidized.

Interestingly, CB~26, which has the lowest $A_V$ and highest flux of interstellar photons, has also the lowest radical content. This is because the higher temperature in CB~26 (see Table~\ref{tab-phys}) allows a greater radical mobility (especially for atomic H) in the mantle and, thence, reactivity. Low radical content in CB~26 ices is characteristic also for other compound classes (sections \ref{r-n}, \ref{r-s}, \ref{r-com}).

Because ice in CB~17~S is initially rich in O$_2$ (figure~\ref{att-o}), it has also a higher abundance for H$_2$O$_2$, which forms via reaction~14 (Table~\ref{tab-reac}). Substantial amounts of H$_2$O$_2$ (relative abundance $X_{\rm H_2O_2}$ in excess of $10^{-6}$) in sublayer~3 for CB~17~S and B~68 are able to form from water photodissociation products that combine together after all CO has been oxidized in that sublayer.

\subsubsection{Comparison with observations}
\begin{figure*}
  \includegraphics[width=18.0cm]{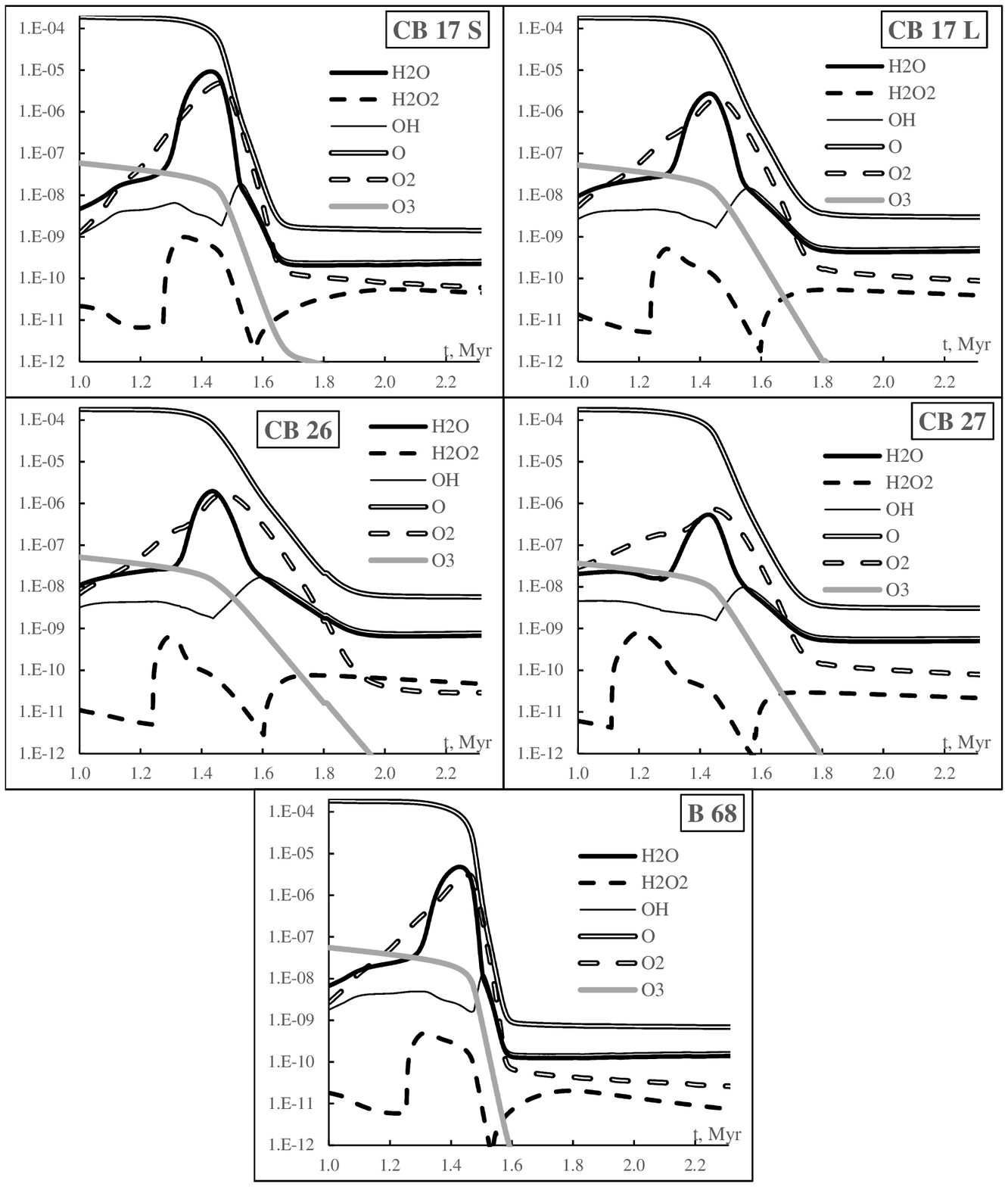}
 \vspace{-7cm}
 \caption{Calculated abundance of selected oxygen species in gas phase, relative to hydrogen.}
 \label{att-o-gas}
\end{figure*}
Figure~\ref{att-o-gas} shows that the calculated abundance of gaseous hydrogen peroxide H$_2$O$_2$ can be reasonably high for a short period, up to $9\times10^{-10}$ relative to total hydrogen atomic number density for CB~17~S. The abundance of its associated O$_2$H radical, however, is lower by three orders of magnitude. Both these species have been observed in interstellar medium with roughly similar abundances of $5\times10^{-11}$ \citep{Bergman11,Parise12}. It has been suggested that reactive desorption is responsible for transferring these species from the surface to the gas \citep{Du12}. However, reactive desorption may be too inefficient to maintain a significant amount of such heavy species in the gas \citepalias{Kalvans15b}. The comparison of abundances here is indicative, only, because of the limits of the 0D model.

Because practically all of H$_2$O$_2$ in ice resides in the sublayers, photodesorption from subsurface ice may play a role in the existence of gas-phase H$_2$O$_2$ and O$_2$H. More often than not, dissociative photodesorption occurs, when molecule fragments are ejected, instead of intact species \citep{Andersson06,Andersson08}. This aspect has not been considered in the present model, or in the model by \citet{Du12}. It likely that photodesorption of H$_2$O$_2$ also induces a high gas-phase abundance of O$_2$H.

Beginning with \citet{Goldsmith02}, molecular oxygen O$_2$ has been repeatedly observed in shocked gas. A few detections have been associated with the cold dense core $\rho$~Ophiuchi A, with inferred abundances in the range $2.5\times10^{-7}...1.25\times10^{-6}$ relative to total hydrogen. Such a high abundances likely can be observed only during a short period in the core's evolution \citep{Larsson07,Liseau10}. These results agree with the calculation results shown in figure~\ref{att-o-gas}, where the maximum abundances lie in the range $7\times10^{-7}...4.5\times10^{-6}$. A quantitative comparison is possible in this particular case because of the centrally localized nature of the observed source \citep{Liseau10}.

\subsection{Nitrogen chemistry in ice}
\label{r-n}

\subsubsection{Chemistry of important species}
\begin{figure*}
 \vspace{-2cm}
  \includegraphics[width=18.0cm]{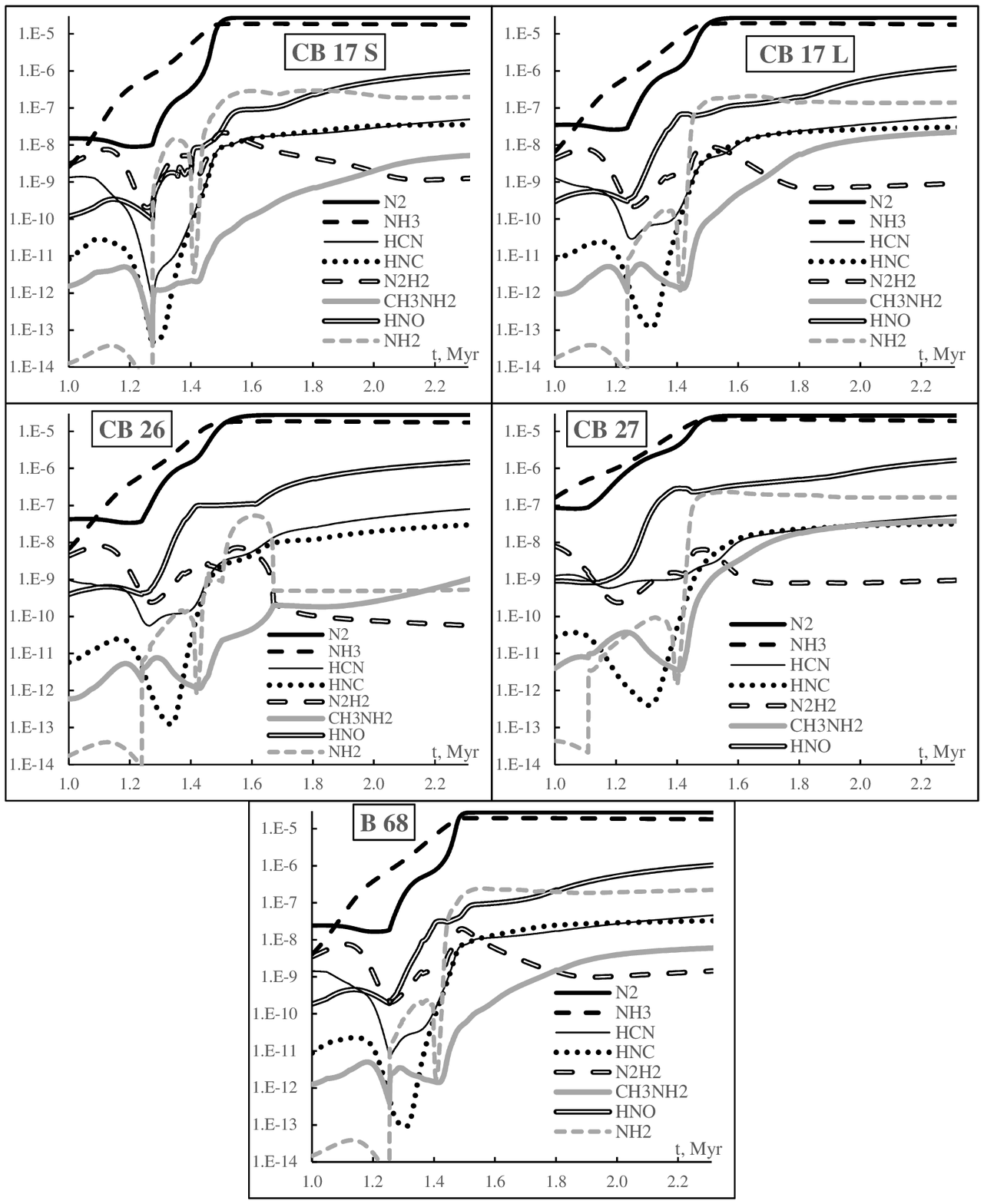}
 \vspace{-7cm}
 \caption{Calculated abundance for important nitrogen species in ice, relative to hydrogen.}
 \label{att-n}
\end{figure*}
\begin{figure*}
 \vspace{-2cm}
  \includegraphics[width=18.0cm]{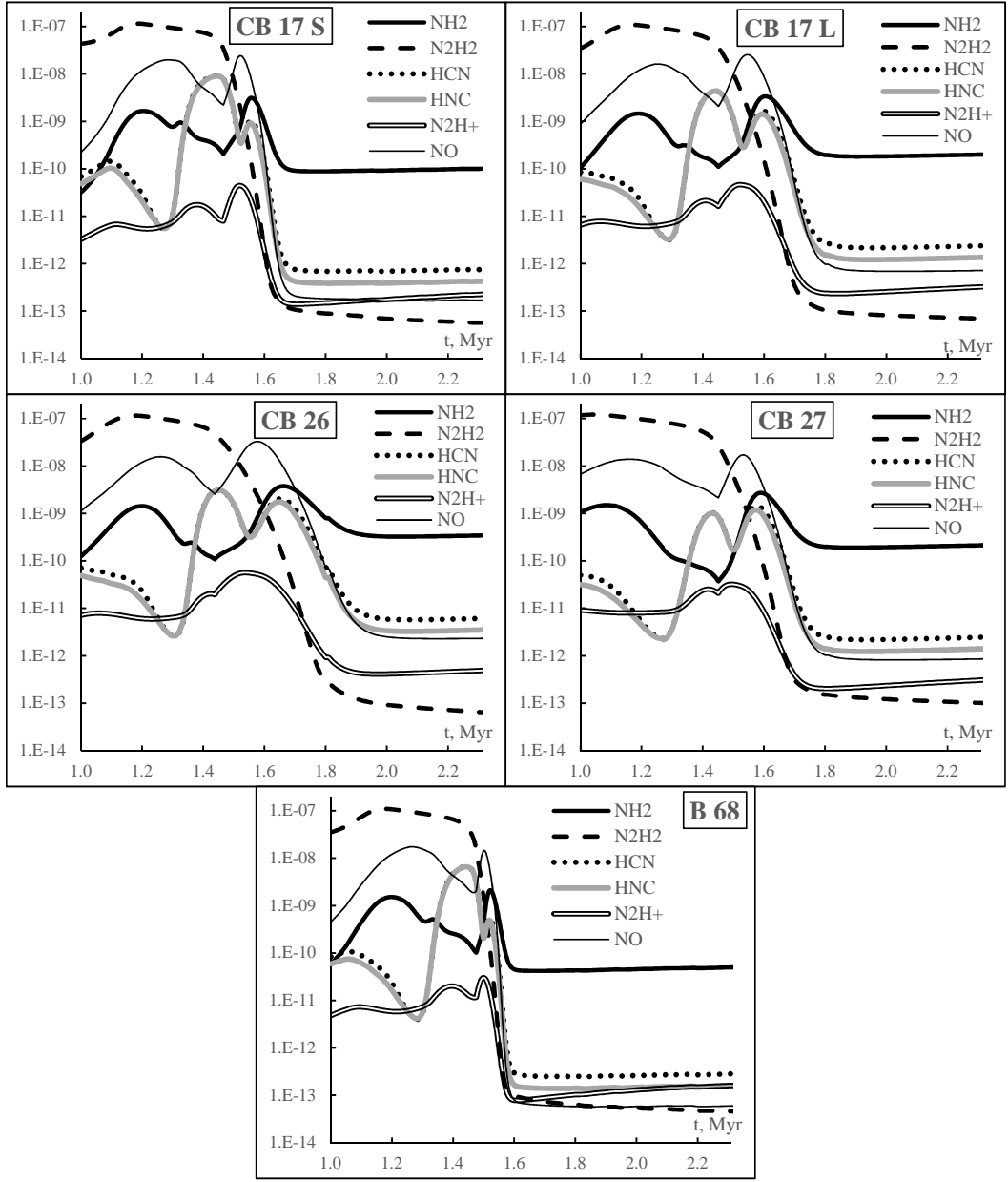}
 \vspace{-7cm}
 \caption{Calculated abundance of important nitrogen species in gas phase, relative to hydrogen.}
 \label{att-n-gas}
\end{figure*}
Figures \ref{att-n} and \ref{att-n-gas} show that ammonia NH$_3$ and molecular nitrogen N$_2$ are the main nitrogen reservoirs in ice and gas. Nitrogen chemistry is regulated by proportions of N$_2$ and NH$_3$ in the surface and the three sublayers (section~\ref{r-gen}).  The relatively clear division of NH$_3$ (in sublayer~3) and N$_2$ (in sublayers 1 and 2) means that N, NH, and NH$_2$ radicals are not equally available in all sublayers. Interesting examples of relatively abundant minor nitrogen species in ice are formamide NH$_2$CHO (discussed in section~\ref{comres}), hydrogen cyanide HCN and isocyanide HNC.

The chemistry of HCN and HNC is controlled by the proportions of CO, N$_2$, and NH$_3$ in each sublayer. HNC is more favored by ice chemistry at earlier times than cyanide HCN, while HCN is the more favored isomer during late times. The two isomers form via different pathways. HNC is mainly synthesized in the reaction NH$_2$+C, while HCN is the product of H addition to CN. The latter forms via C+N. This means that HNC is a product of NH$_3$ and CO mixture photoprocessing, while HCN is synthesized in a N$_2$ and CO mixture. The HNC pathway is more efficient because it requires no subsequent hydrogenation. The result is that 60 (for CB~17~S) to 90\% (for CB~26) of both isomers are concentrated in the outer sublayer~1, where CO and N$_2$ are highly abundant, and NH$_3$ is available, too. The remaining 10-40\% reside in sublayer~2, while sublayer~3 is very poor on HCN and HNC ($<$2\%), because of a lack of CO there. Because the surface closely interacts with gas-phase CO and N$_2$, HCN is the main isomer for the top ice layer.

\subsubsection{Comparison with observations}
\begin{figure*}
 \vspace{-1cm}
  \includegraphics[width=18.0cm]{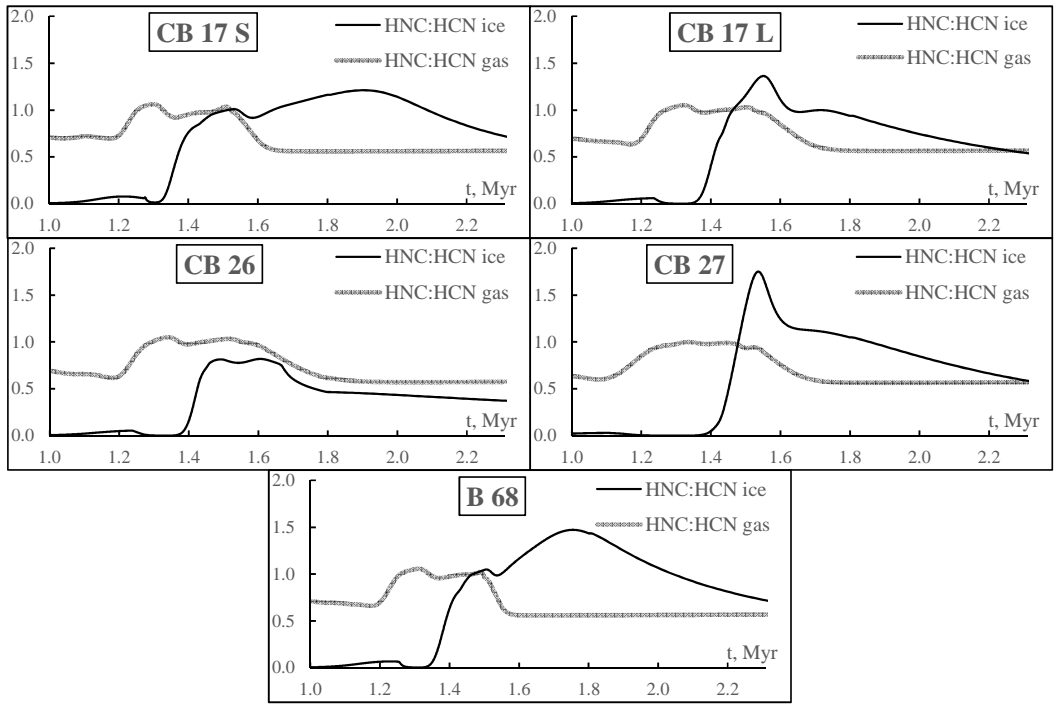}
 \vspace{-14cm}
 \caption{Calculated HNC:HCN abundance ratio in gas and ice.}
 \label{att-hcn}
\end{figure*}
Recent observed NH$_3$ gas-phase abundances, lie in the range $9\times10^{-9}...7.5\times10^{-8}$ relative to total H \citep{Ruoskanen11,Levshakov14}. \citet{Crapsi07} derives a central NH$_3$ abundance of $4.5\times10^{-9}$ for the L~1544 dense core. These values qualitatively agree with calculation results for cores with ages in the range $t=1.5...1.8$Myr.

Figure~\ref{att-hcn} shows the calculated HNC:HCN abundance ratio the center of the respective dark cores in gas and solid phases. We note that the observed gas-phase values of this ratio are in the range of 0.54-4.5, with no distinction between YSOs and starless cores \citep{Hirota98}. HNC:HCN ratio in ice may exceed unity for all cores, except for CB~26, where lower abundance of CO prevents the build-up of HNC. The highest calculated HNC:HCN ratio is in CB~27 ice phase, where it reaches 1.75 briefly at 1.54Myr. Such a peak arises because of effective HNC production in sublayer~1, not observed in other simulations. In turn, this occurs because sublayer~1 in CB~27 is richer in NH$_3$ by a factor of $>$2 than in other models. Finally, NH$_3$ is more abundant because ices in CB~27 are accumulated at lower extinctions, which means a higher proportion of species that form from free atoms on the surface (H$_2$O, NH$_3$, CH$_4$, etc.).

\subsection{Sulfur chemistry in ice}
\label{r-s}
\begin{figure*}
  \includegraphics[width=18.0cm]{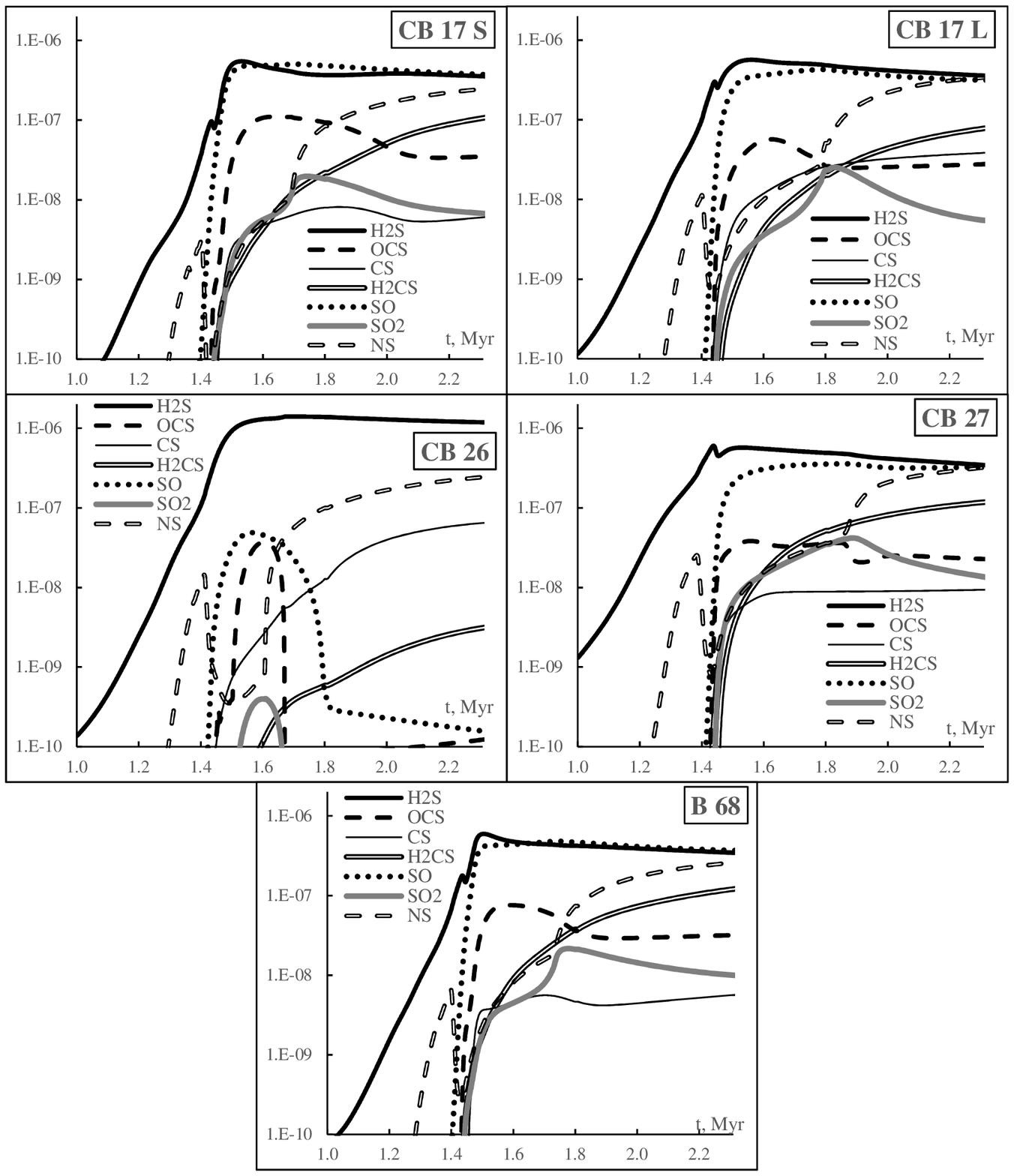}
 \vspace{-8cm}
 \caption{Calculated abundance of important sulfur species in ice, relative to hydrogen.}
 \label{att-s}
\end{figure*}
\begin{figure*}
  \includegraphics[width=18.0cm]{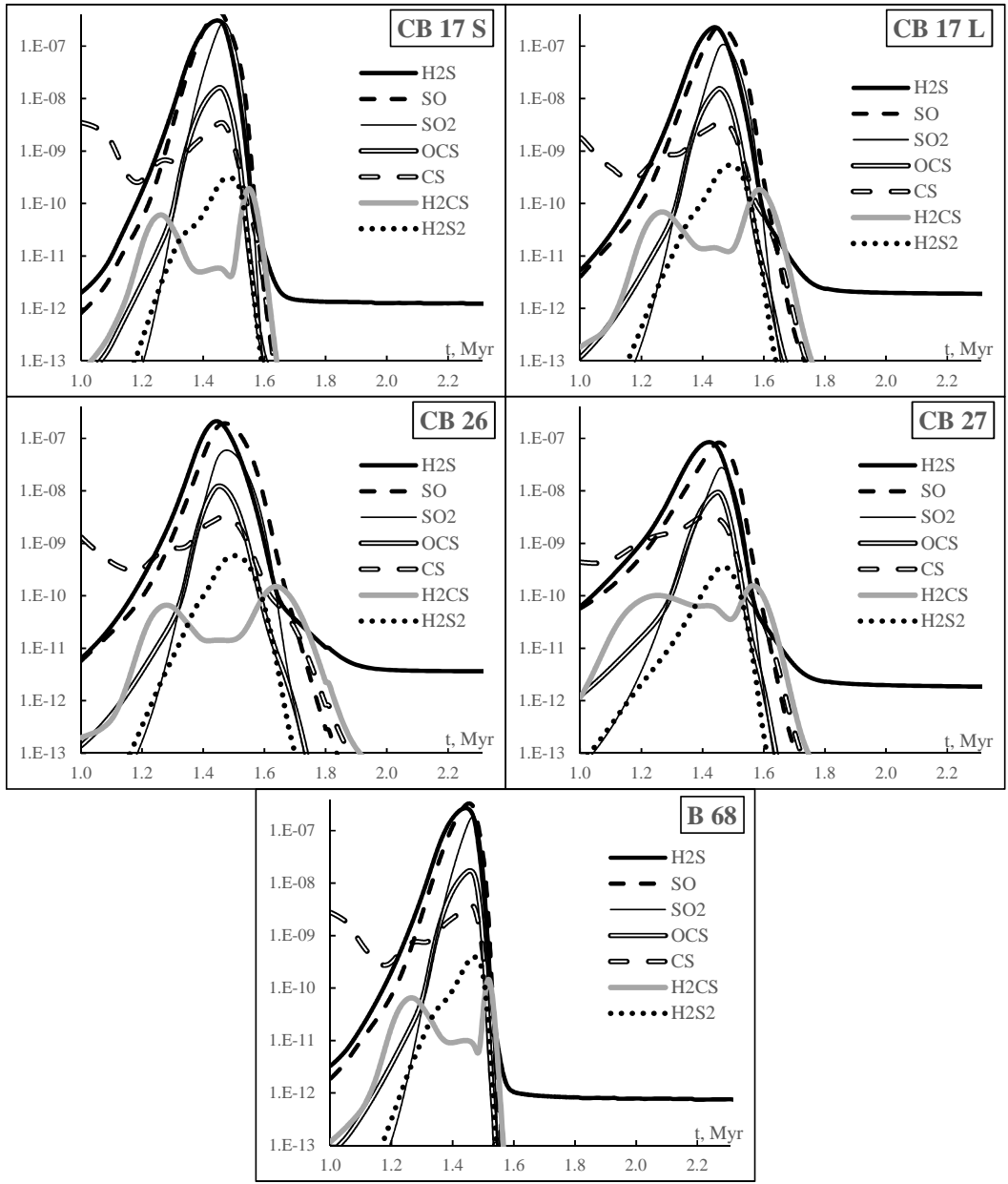}
 \vspace{-7cm}
 \caption{Calculated abundance of important sulfur species in gas phase, relative to hydrogen.}
 \label{att-s-gas}
\end{figure*}
\begin{figure*}
  \includegraphics[width=18.0cm]{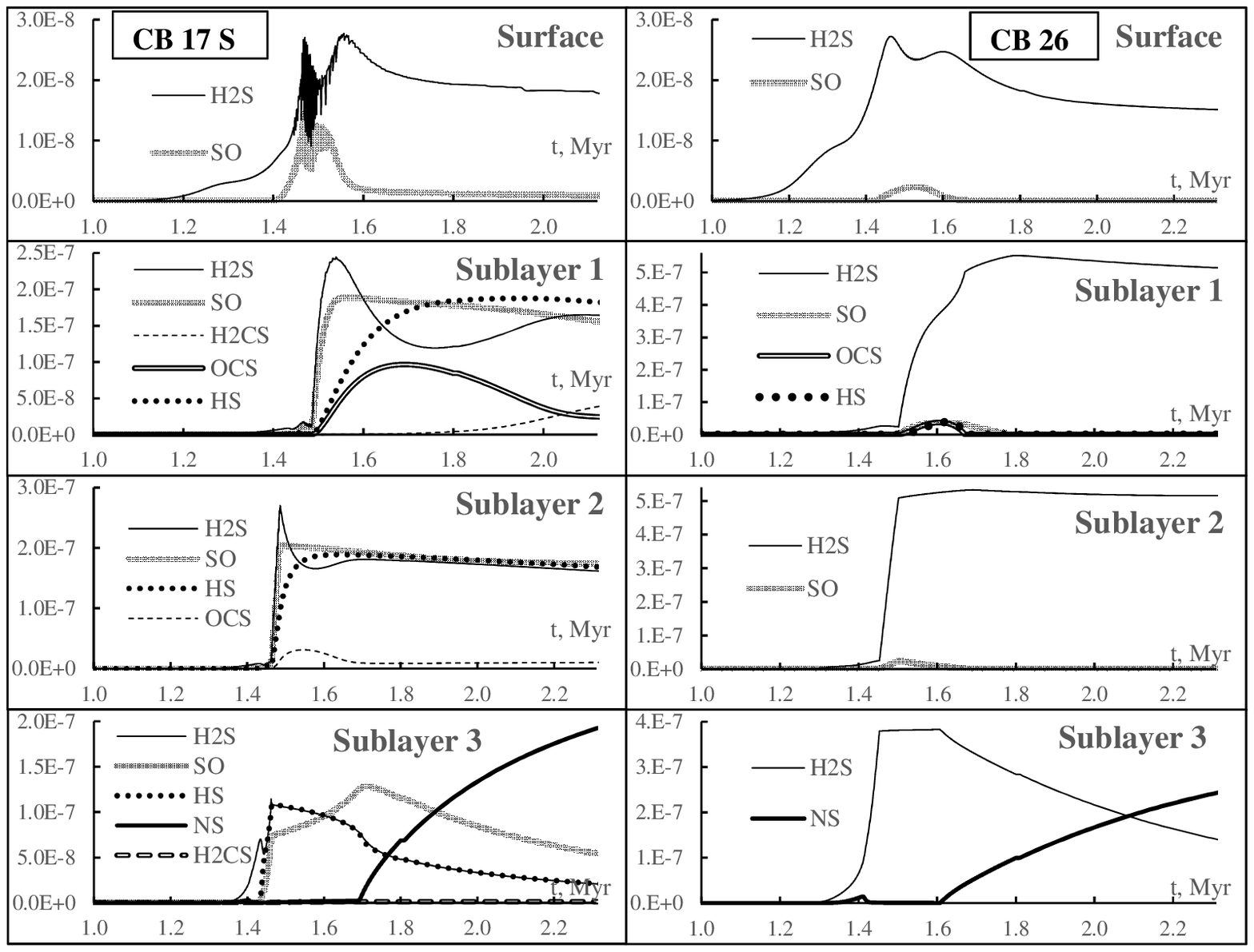}
 \vspace{-13cm}
 \caption{Calculated abundance, relative to hydrogen, of main sulfur species in ice mantle sublayers for CB~17~S and CB~26. The latter represents an significantly photoprocessed ice.}
 \label{att-s-sub}
\end{figure*}
Figures \ref{att-s} and \ref{att-s-gas} show that H$_2$S and SO are the most abundant sulfur species in ice and gas. H$_2$S is more abundant than SO for most of the time. An exception is the CB~26 core, where H$_2$S is the only dominant species and SO ice forms with a maximum relative abundance of $5\times10^{-8}$ at 1.56Myr, only, thanks to surface reactions. Then, SO is quickly consumed by ice mantle reactions because of the higher temperature in the CB~26 core. In the other four cores, the HS radical accumulates and has time to react with atomic O, producing SO. Increased mobility of hydrogen in CB~26 ice does not allow the accumulation of HS, S, and many other radicals. H$_2$S can be expected to be the main sulfur reservoir for photon-dominated ices. Notably, CB~26 still has a maximum SO gas-phase abundance comparable to that in other modeled cores (figure~\ref{att-s-gas}).

The abundances of minor sulfur molecules are affected by the proportions of major ice constituents. Molecules, whose synthesis is associated with CO and its daughter radicals (primarily atomic C, atomic O is available also from CO$_2$), include OCS, CS, H$_2$CS, and C$_2$S. These are primarily associated with sublayers 1 and 2 (see figure~\ref{att-s-sub}). Oxidized forms SO$_2$ and NS arise in sublayer~3. The above is true for cores CB~17~L, CB~17~S, CB~27, and B~68. For CB~26, the diversity is largely limited by the dominating H$_2$S. The efficient synthesis on the surface and subsequent reactive desorption explain why H$_2$S is the most abundant sulfur gas phase molecule for all modeled cases.

Modeling results show that H$_2$S and SO ices are produced during the freeze out epoch and contain most of the sulfur reservoir in interstellar ices. The reaction set for sulfur species is rather poor and does not include sulfurous and sulfuric acids (H$_2$SO$_3$ and H$_2$SO$_4$, respectively) that likely are very stable forms of sulfur in watery environment. Oxoacids and their derivatives are possible and non-detectable forms of sulfur in star-formation regions \citep{Kalvans10}. Given the rich chemistry of sulfur and the radical nature of the SO molecule, it is unlikely that SO is a final, stable, and abundant form of S in interstellar or circumstellar ices. More likely, in terms of this model, it can be viewed as a representation of oxidized sulfur \citep[see also][]{Scappini03}.

Figure~\ref{att-s-sub} shows an example of depth-dependent composition for sulfur species. H$_2$S and SO are dominant at all depths, except for sublayer~3 at late times, where the abundance of NS is highly enhanced. Although this is probably because of the limited reaction network for sulfur, this illustrates the oxidation of S during longer timescales. Oxidation products are SO, SO$_2$, NS or perhaps, other species, not included in the network. Sulfur in the outer sublayer~1 is less susceptible to oxidation thanks to a higher availability of atomic H.

The high abundance of NS in sublayer~3 for CB~17~L, CB~17~S, CB~27, and B~68 is an example how local conditions in the ice mantle may give rise to peculiar chemical features. NS is produced in all layers via the reactions N~+~S and NH~+~S. In sublayers 1 and 2 it is also efficiently destroyed via reactions with atomic C and N. Sublayer~3 is poor with CO and N$_2$. Thus, C and N are unavailable, and NS accumulates after all remaining CO has been converted into CO$_2$.

\section{Results: complex organic molecules}
\label{comres}

When compared to observational evidence, the synthesis of organic species in the model can be characterized as moderately efficient. For some species, gas phase abundances are in temporal qualitative agreement with the observations, for others even ice abundances are well below expectations. To improve this situation, we introduce two simple changes in the model -- a modification to CO and H$_2$CO hydrogenation activation energies and, subsequently, a mild and short temperature spike during the quiescent phase. The results of the three simulations are described in the following sections.

To provide an indication of the observed abundances of gas-phase COMs relative to the calculated values, the observed values have been marked in their corresponding figure panels. The observed values have no temporal or spatial constraints with regard of the present model. The 0D character of the simulation means that a qualitative comparison is possible, only. We assume that the synthesis of an organic molecule is reproduced with a degree of success, if its calculated ice abundance is not lower than its observed gas-phase abundance.

\subsection{COMs: Standard model}
\label{r-com}

\subsubsection{The chemistry of COMs}
%
\begin{figure*}
 \vspace{-2.5cm}
  \includegraphics[width=18.0cm]{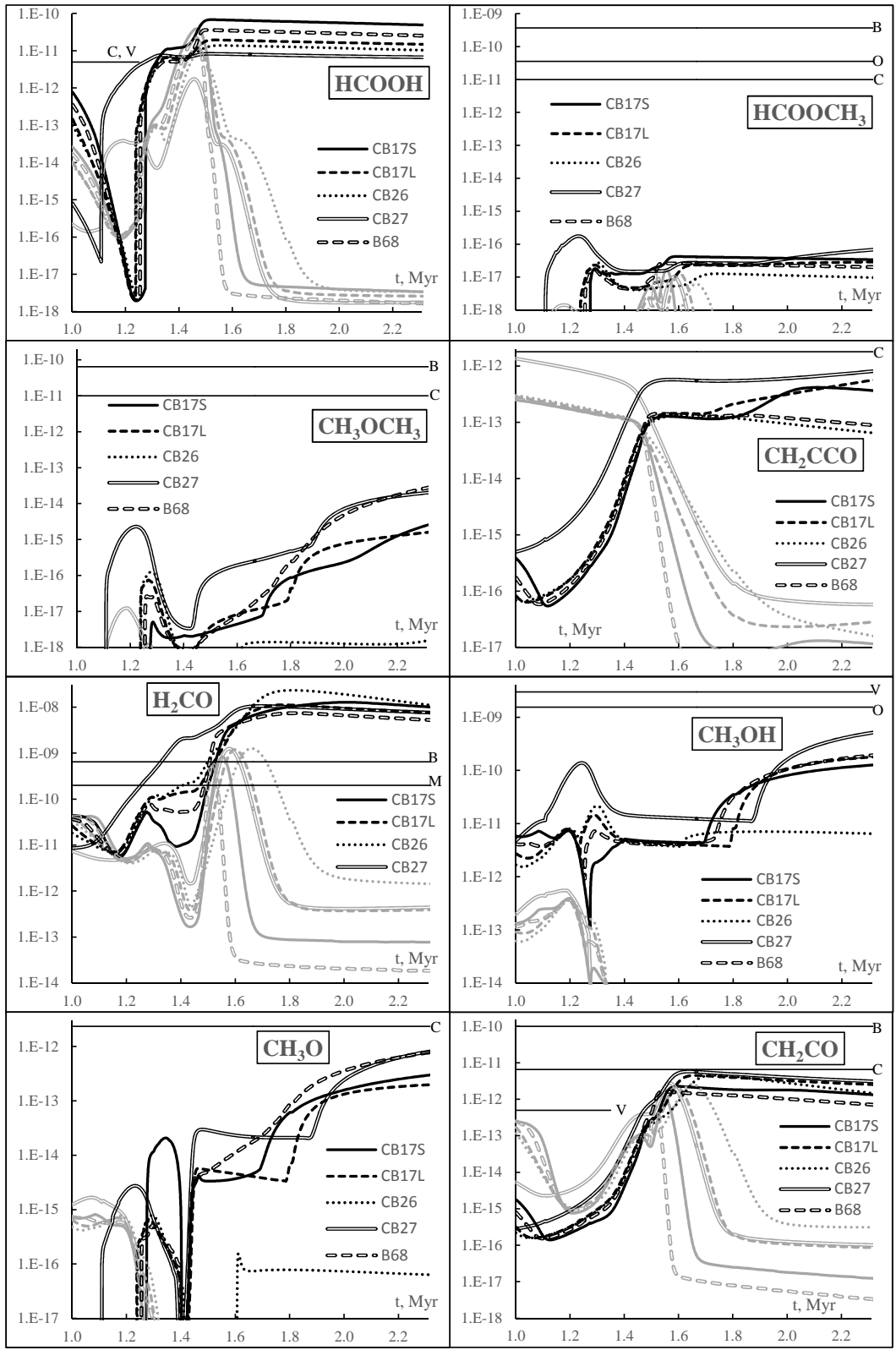}
 \vspace{-2.5cm}
 \caption{Calculated abundances, relative to hydrogen, of observed organic molecules, set No.~1. Gray lines: gas-phase abundance; black lines: abundance in ice (surface and three mantle sublayers). Contraction (Phase~1) ends and the stable core Phase~2 begins at times between 1.4 and 1.5Myr for all simulations. References: M -- \citet{Marcelino05}; O -- \citet{Oberg10}; B -- \citet{Bacmann12}; C -- \citet{Cernicharo12}; V -- \citet{Vastel14}.}
 \label{att-org1}
\end{figure*}
\begin{figure*}
 \vspace{-2cm}
  \includegraphics[width=18.0cm]{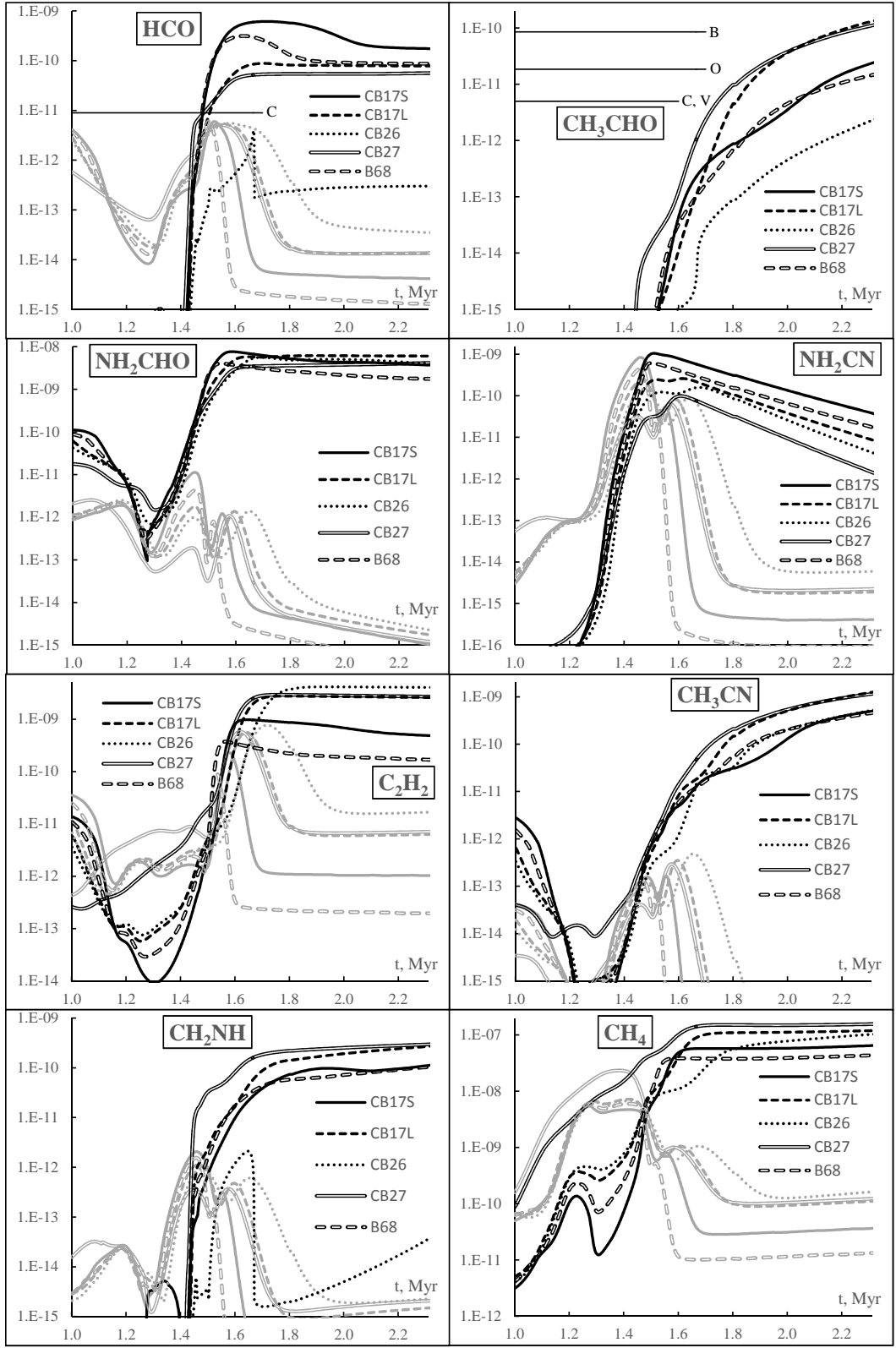}
 \vspace{-3cm}
 \caption{Calculated abundances, relative to H, of observed and other selected organic molecules, set No.~2. References as in figure~\ref{att-org1}.}
 \label{att-org2}
\end{figure*}
\begin{figure*}
  \includegraphics[width=18.0cm]{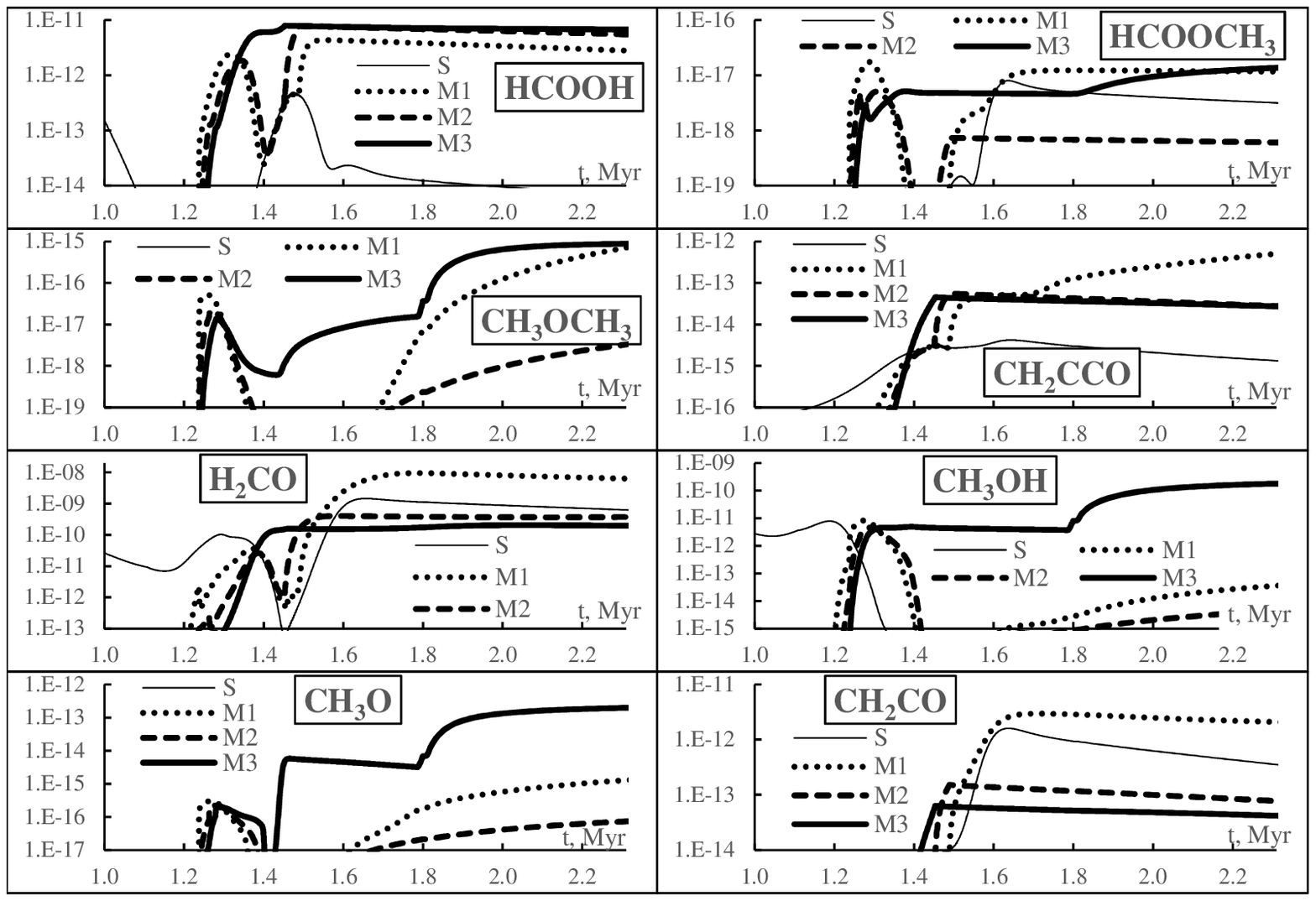}
 \vspace{-15cm}
 \caption{Calculated abundances, relative to H, in ice sublayers for selected organic molecules for the CB~17~L core, set No.~1.}
 \label{att-org-sub1}
\end{figure*}
\begin{figure*}
  \includegraphics[width=18.0cm]{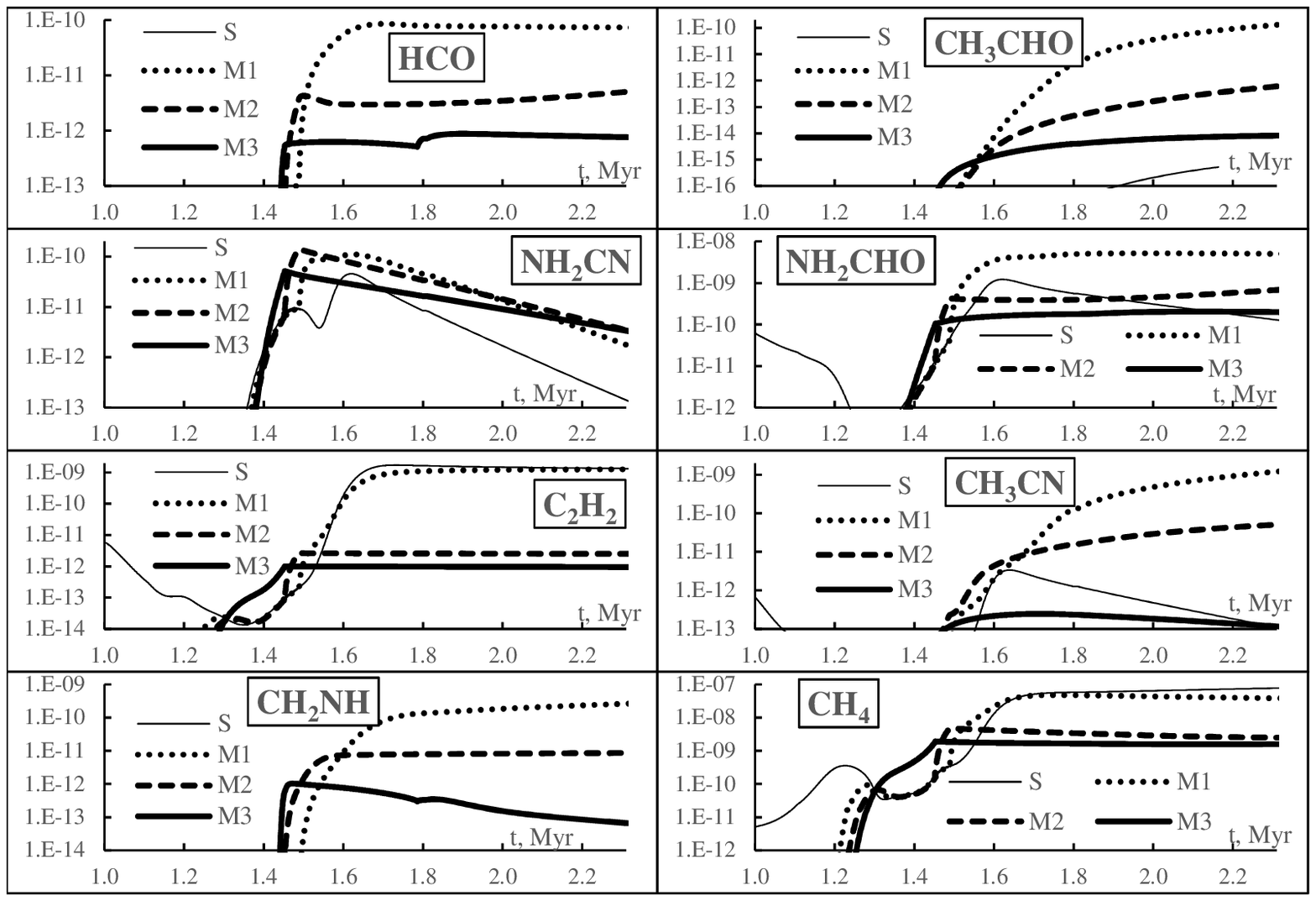}
 \vspace{-15cm}
 \caption{Calculated abundances, relative to H, in ice sublayers for selected organic molecules for the CB~17~L core, set No.~2.}
 \label{att-org-sub2}
\end{figure*}
The "Standard" model here is the one described in full in section~\ref{meth}. Figures \ref{att-org1} and \ref{att-org2} show the calculated abundances in the center of the starless cores, along with known observational values. The species can be categorized in several overlapping groups:
\begin{itemize}
\item those synthesized via the HCO radical, e.g., formic acid HCOOH, formaldehyde H$_2$CO, methyl formate HCOOCH$_3$, formamide NH$_2$CHO, and acetaldehyde CH$_3$CHO;
\item those synthesized via the CH$_3$O radical, e.g., methanol; CH$_3$OH, dimethly ether CH$_3$OCH$_3$, HCOOCH$_3$, and as a subgroup, species whose synthesis involves CH$_3$, which is a daughter of methanol. This subgroup includes CH$_3$CHO, acetonitrile CH$_3$CN, and methane CH$_4$;
\item carbon chain molecules, e.g., CH$_3$CHO, propynone CH$_2$CCO, ketene CH$_2$CO, and acetylene C$_2$H$_2$;
\item species that are typically formed via atom addition on the surface, e.g., CH$_4$, H$_2$CO, HCO, HCOOH CH$_2$CCO, CH$_2$CO, C$_2$H$_2$, methyl imine CH$_2$NH, cyanamide NH$_2$CN, and CH$_3$CN. A relatively high gas-phase abundance during the freeze-out epoch is characteristic for these species.
\end{itemize}

The hydrogenation of CO and H$_2$CO are the two most important steps that regulate the general production efficiency for many oxygen-containing COMs in bulk ice. Figures \ref{att-org-sub1} and \ref{att-org-sub2} show the relative abundances of organic species in the four ice layers considered in the models for CB~17~L. This core can be regarded as a `median' case among the five models, without the extremes in ice composition shown by CB~17~S and B~68 on one hand, and CB~26 and CB~27 on the other hand (section~\ref{r-mform}).

Most of the species with abundances shown in figures \ref{att-org-sub1} and \ref{att-org-sub2} have their peak abundances highest in the outermost sublayer~1. This indicates that even a partial ice sublimation might be sufficient to reproduce the observed gas-phase abundances of many COMs. In sublayer~1, CO is available in vast quantities at any integration time points, while the necessary H atoms are produced in H$_2$O or H$_2$ photodissociation. With the present 0D model it is not possible to determine the percentage of ice mass that should be sublimated. If the modeled ices are assumed to be representative, this proportion would be $\approx1-10$\%.

\subsubsection{Comparison with observations}

Figures \ref{att-org1} and \ref{att-org2} show that the model predicts relatively high gas-phase abundance for some species in the HCO group -- formic acid HCOOH, formaldehyde H$_2$CO, and formamide NH$_2$CHO. Although the observed gas-phase abundances are reached for a short period of time, this occurs during the infall Phase~1 and likely cannot be treated as an agreement with observations of stable starless cores.

For the CH$_3$O group, the calculated abundances are significantly lower than those observed in interstellar conditions. Even assuming instant ice sublimation, observed gas-phase abundances cannot be reproduced for methyl formate HCOOCH$_3$, dimethyl ether CH$_3$OCH$_3$, and methanol CH$_3$OH. This result is similar to that of earlier prestellar core models without \citep{Garrod06,Garrod08} and with bulk ice photochemistry \citep[][\citetalias{Kalvans15b}]{Garrod13a}. It is this discrepancy that prompts us to proceed with a more detailed investigation on the chemistry of COMs.

\subsection{The Compromise model}
\label{mcompr}

\subsubsection{Changes relative to the Standard model}
\begin{table*}
\begin{center}
\caption{Changes in reaction data for the compromise model.}
\label{tab-ea}
\begin{tabular}{clccc}
\tableline\tableline
 &  & \multicolumn{3}{c}{Activation barrier $E_A$, K} \\
No. & Reaction & Standard\tablenotemark{a} & \citet{Fuchs09} & Compromise \\
\tableline
17 & $\rm H + CO \longrightarrow HCO$ & 2500 & 390-520 & 1600 \\
18 & $\rm H + H_2CO \longrightarrow CH_3O$ & 2100 & 415-490 & 415 \\
19 & $\rm H + H_2CO \longrightarrow CH_2OH$ & 2500 & 415-490 & 415 \\
\tableline
\end{tabular}
\tablenotetext{a}{The initial values used in this paper}
\end{center}
\end{table*}
In the present surface reaction network \citep{Garrod08} the activation energy barriers $E_A$ for reactions H+CO and H+H$_2$CO are higher than 2000K, which means that they are inefficient for cold core conditions. Experimental evidence points to much lower values for $E_A$, 300-600K \citep{Awad05,Fuchs09}. However, putting such low $E_A$ values into the present model results in that almost all carbon in ice is in the form of methanol, which is unrealistic. Higher $E_A$ is especially required for the CO+H reaction.

We present results of model calculations that use the experimental, low $E_A$ value for formaldehyde hydrogenation. For CO hydrogenation $E_A$, an empirical compromise value was used, which lies between the low experimentally detected $E_A$ of \citet{Fuchs09} and the high (2500K) value proposed by \citet{Garrod08}. Activation energies for the three important CO hydrogenation reactions have been summarized in Table~\ref{tab-ea}. We dub this the "Compromise model" for further reference.

\subsubsection{Comparison with observations}
\label{r-compr}
\begin{figure*}
 \vspace{-2cm}
  \includegraphics[width=18.0cm]{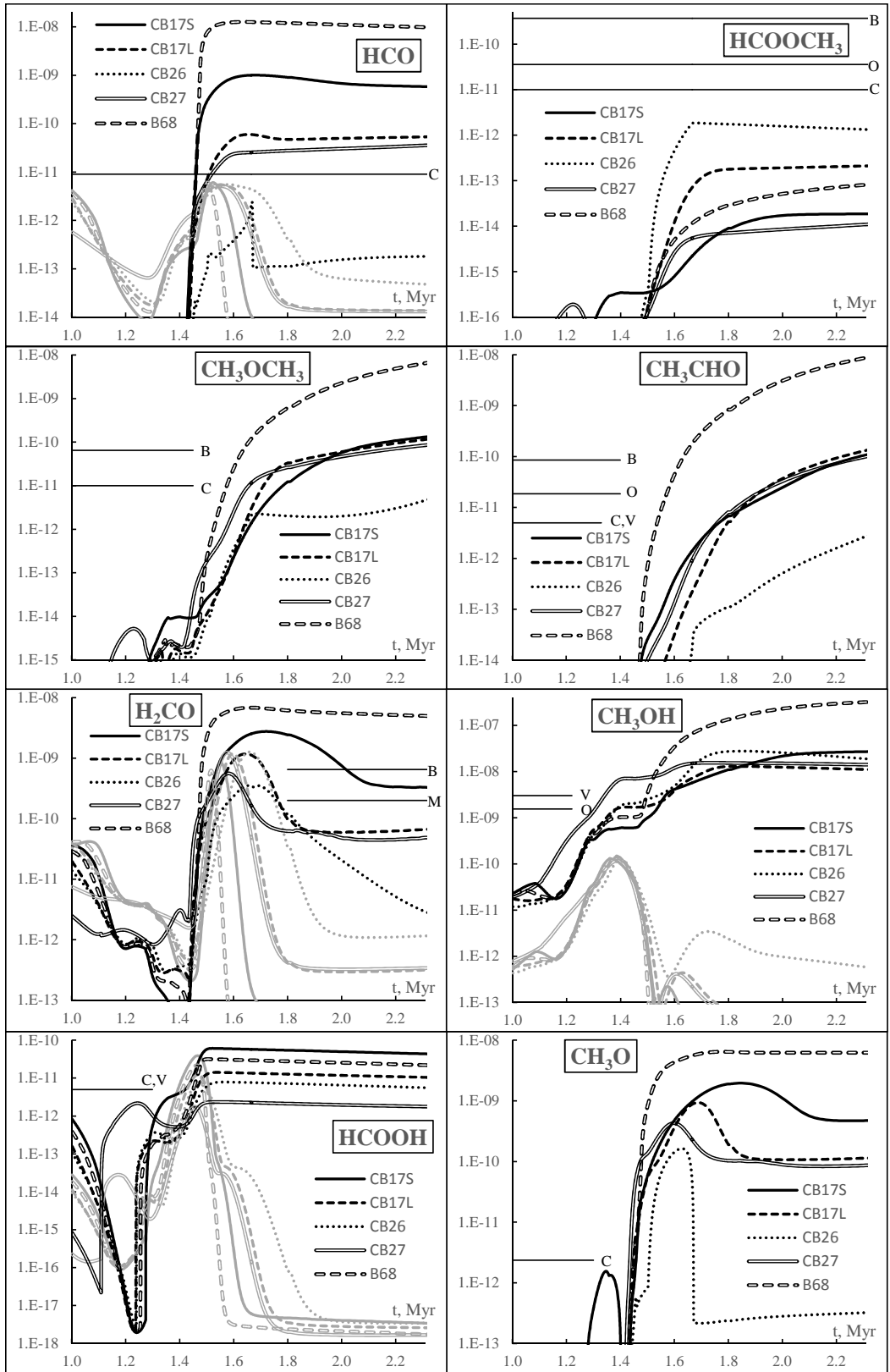}
 \vspace{-3cm}
 \caption{Calculated abundances, relative to H, of selected organic molecules for the Compromise model. Gray lines: gas-phase abundance; black lines: abundance in ice (surface and three mantle sublayers). For references see the caption of figure~\ref{att-org1}.}
 \label{att-compr}
\end{figure*}
Figure \ref{att-compr} shows species, whose abundance is substantially affected with the changes in activation energies for reactions 17, 18, and 19 in Table~\ref{tab-ea}. Comparison with the Standard model results (figures \ref{att-org1} and \ref{att-org2}) reveals that the ice abundance of species that are formed via CH$_3$O is increased by up to six orders of magnitude (CH$_3$OCH$_3$ in B~68). Importantly, the ice abundances of CH$_3$OCH$_3$, CH$_3$OH, and CH$_3$O are now above or similar to the detected gas phase values.

The synthesis of methyl formate HCOOCH$_3$ depends on both HCO and CH$_3$O radicals. We found it impossible to achieve a sufficiently high ice abundance for HCOOCH$_3$ with changes in $E_A$ for reactions 17, 18, and 19. Thus, the calculated relative ice abundance for this molecule remains approximately four orders of magnitude below the observed gas-phase values, which are between $10^{-11}$ and $10^{-9}$ \citep{Oberg10,Bacmann12,Cernicharo12}.

Methyl formate is also inefficiently produced in several other models considering either starless or prestellar cores \citep{Garrod13a,Vasyunin13a,Vasyunin13b}. \citet{Chang14} are able to efficiently synthesize HCOOCH$_3$ with a Monte-Carlo model, but their results also show a likely excess of CH$_4$ and a shortage of CO$_2$ -- discrepancies generally not observed in the abovementioned other models. These results may indicate that the inefficient synthesis of HCOOCH$_3$ is not an artifact, caused by the limits of the present study. A higher temperature may be necessary for an efficient synthesis of HCOOCH$_3$, as suggested by \citet{Garrod06}.

\subsubsection{Chemistry differences with the Standard model}
\begin{figure}
 \vspace{-1cm}
  \hspace{-1cm}
  \includegraphics[width=16.0cm]{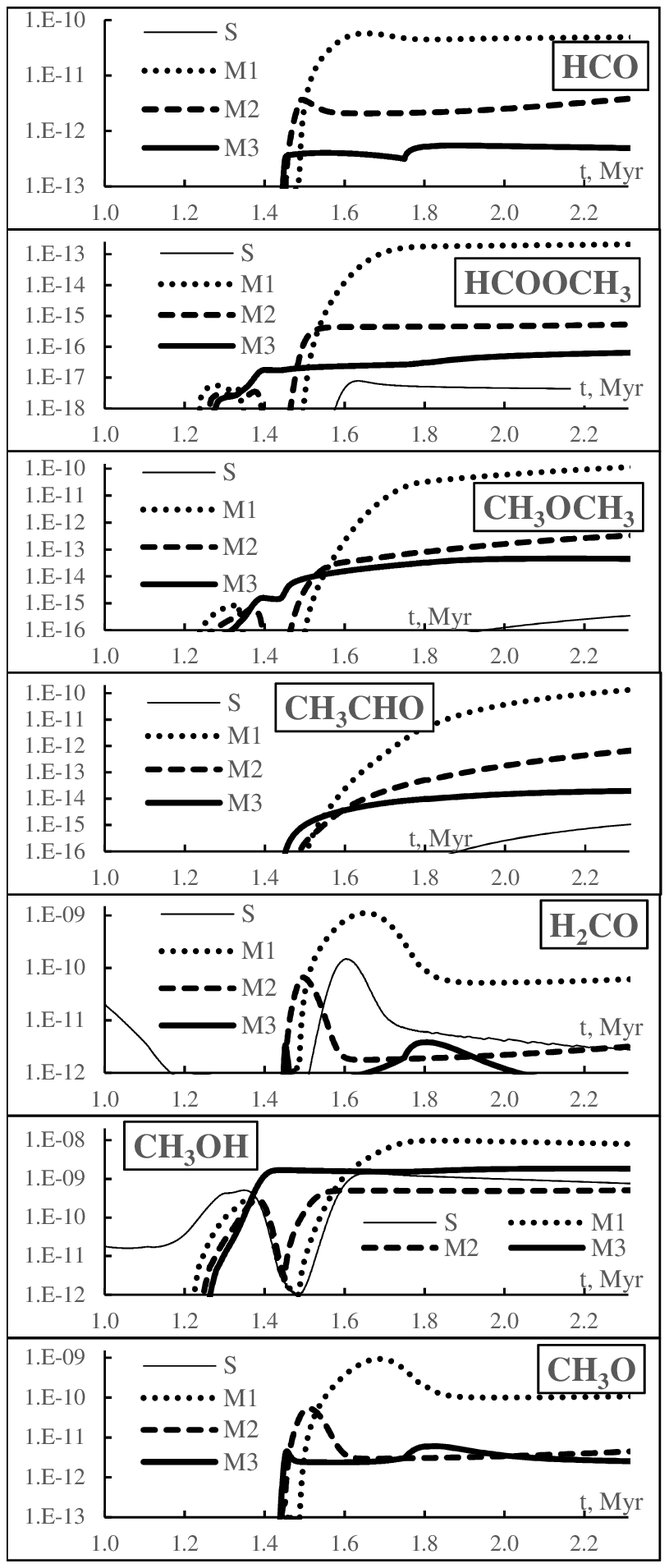}
 \vspace{-7cm}
 \caption{Calculated abundances, relative to H, in ice sublayers for selected organic molecules, Compromise model, CB~17~L core.}
 \label{att-compr-sub}
\end{figure}
The most significant abundance increase for the CH$_3$O group is observed in the B~68 model. First, the lower temperatures in this core mean a low mobility for atomic H in ice. Second, the abundant CO is now more easily hydrogenated, thanks to the lower $E_A$ and CO soaks up much of the available atomic hydrogen in the mantle. These two factors result an accumulation of radicals so that they have greater chance to react with other multi-atom species. For the present discussion on COMs, the most important radicals are CH$_3$O and HCO. The effect is especially visible for B~68 and CB~17~S.

For species that are formed via HCO, the situation is less clear because formaldehyde H$_2$CO is consumed by reactions 18 and 19. In cores CB~17~L and CB~27, formaldehyde has its maximum ice abundance decreased below the detected gas-phase values. A similar situation can be observed for formic acid HCOOH, directly related to HCO.

Figure~\ref{att-compr-sub} shows that in the compromise model the abundances of COMs are increased in all sublayers (cf. figures \ref{att-org-sub1} and \ref{att-org-sub2}). The main part of their synthesis still occurs in the outer sublayer~1. Species whose synthesis involve CH$_3$O and CH$_3$, do not reach a steady abundance until the end of the simulation, mainly because they can be regarded as the end products of CO hydrogenation sequence.

\subsection{The synthesis of COMs in ice in a temporal warm-up event}
\label{r-warm}
\begin{figure*}
 \vspace{-2.5cm}
  \includegraphics[width=18.0cm]{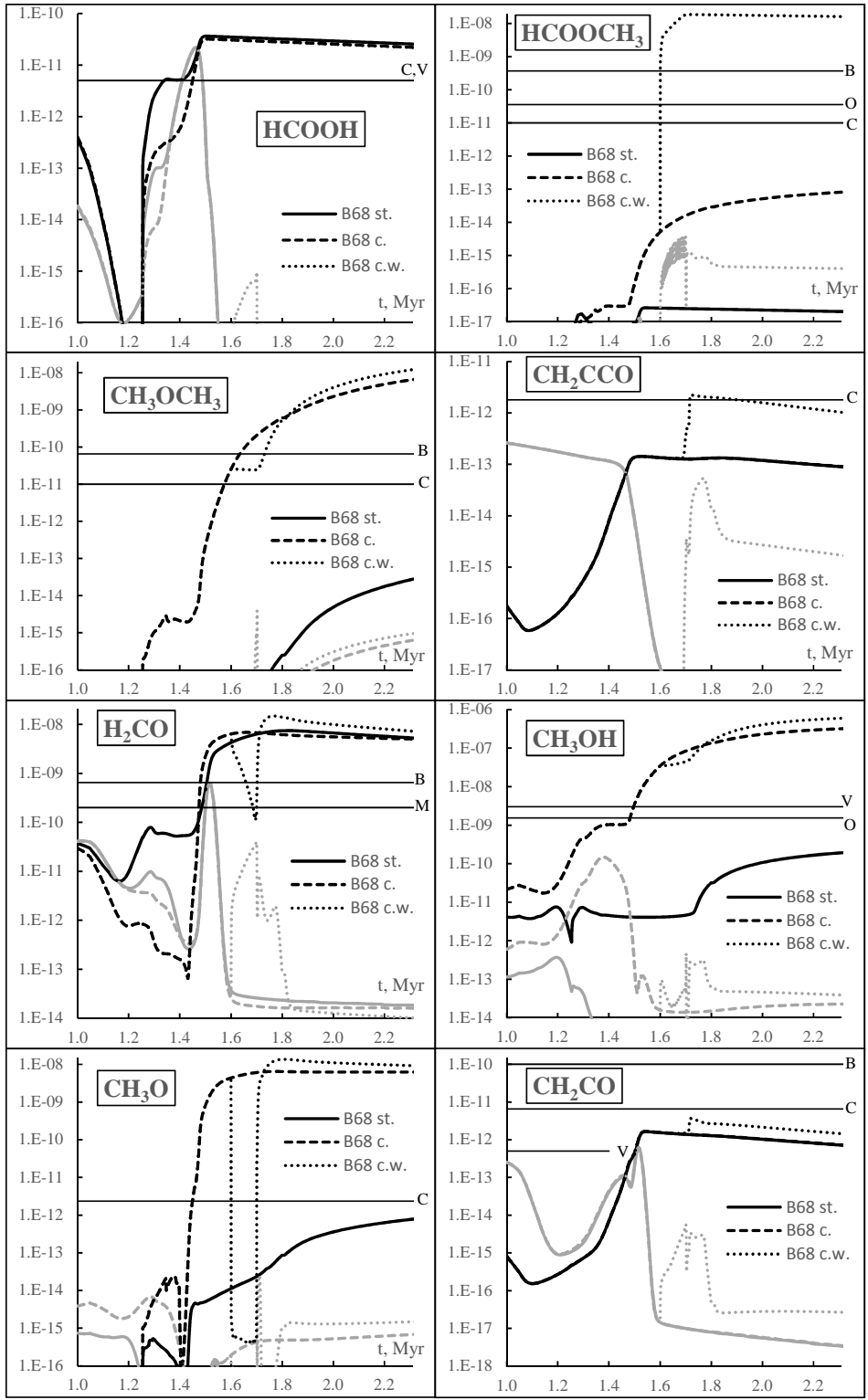}
 \vspace{-4cm}
 \caption{Comparison of calculated COM abundances, relative to H, for the B~68 core between the Standard model (st.), Compromise model (c.), and Compromise model with a warm temperature spike (c.w.), set No.~1. The two latter models begin to show differences only after 1.6Myr, when the temperature spike occurs. References as in figure~\ref{att-org1}.}
 \label{att-compr1}
\end{figure*}
\begin{figure*}
 \vspace{-2cm}
  \includegraphics[width=18.0cm]{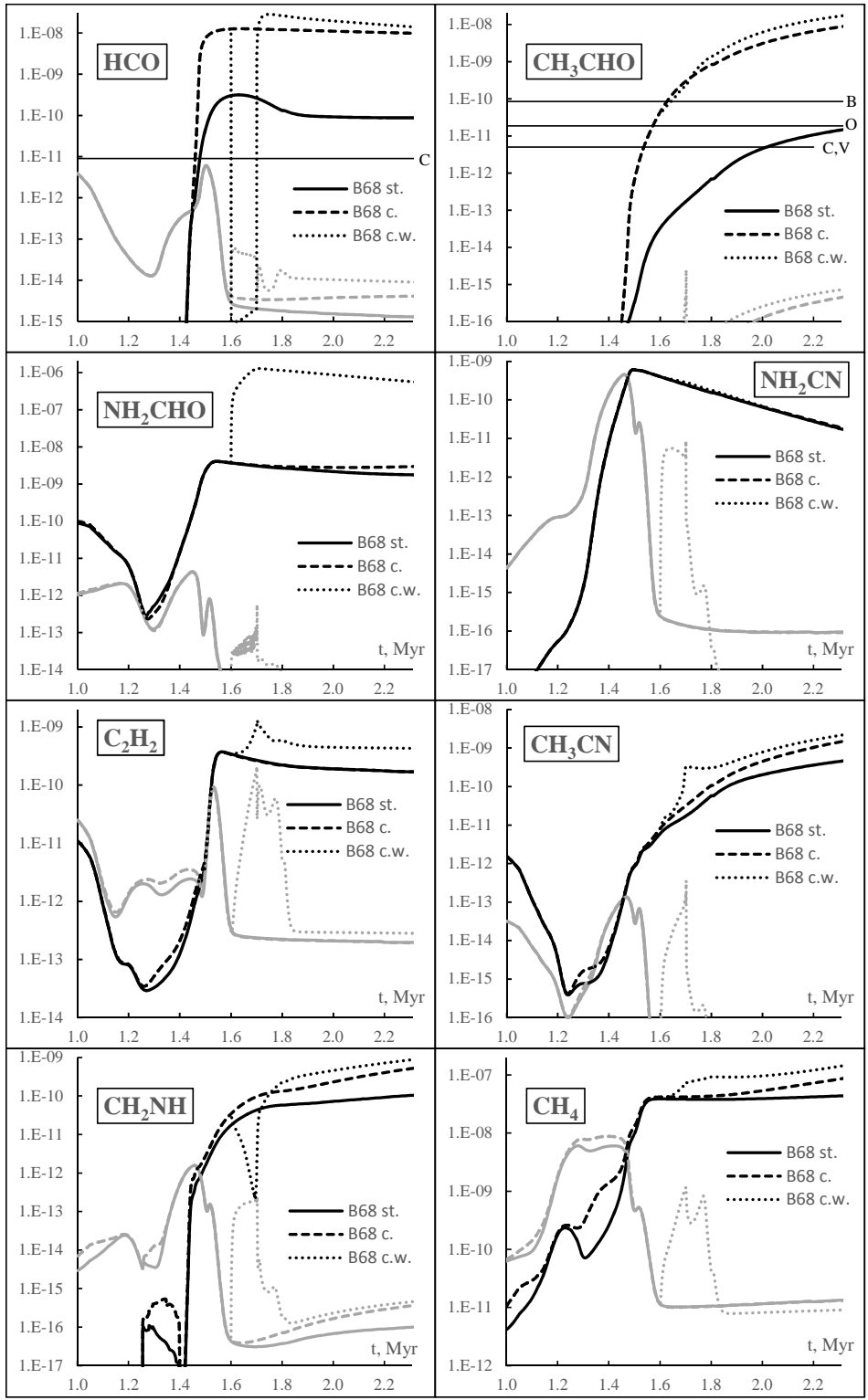}
 \vspace{-4cm}
 \caption{Comparison of calculated COM abundances, relative to H, for the B~68 core between the standard model (st.), compromise model (c.), and compromise model with a warm temperature spike (c.w.) at 1.6Myr, set No.~2. References as in figure~\ref{att-org1}.}
 \label{att-compr2}
\end{figure*}
\begin{figure}
 \vspace{-2cm}
  \hspace{-1.5cm}
  \includegraphics[width=18.0cm]{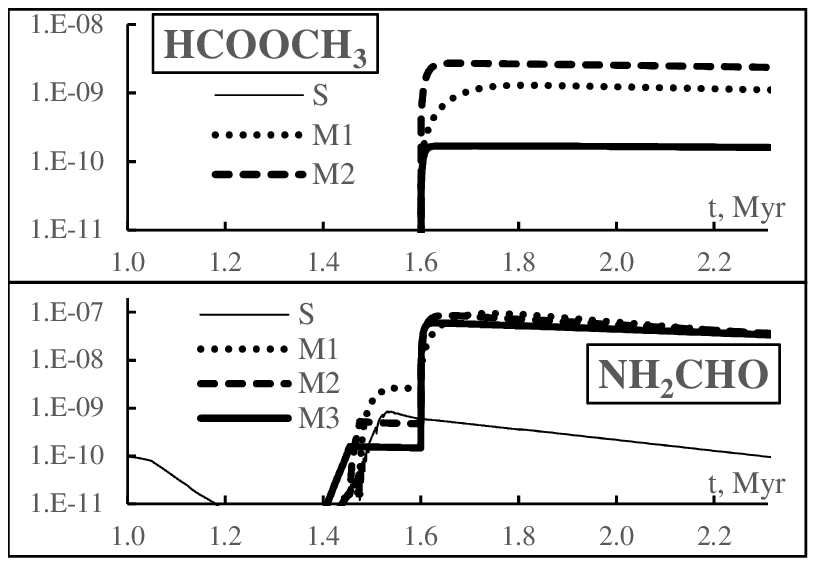}
 \vspace{-19.5cm}
 \caption{Calculated abundances, relative to H, in ice sublayers for species most affected by a 1kyr 20K temperature spike in the Compromise model for the B~68 core.}
 \label{add-sho}
\end{figure}
The discussion in the above section~\ref{r-compr} shows that for some COMs (notably, methyl formate) the calculated abundances are nowhere near their observed values. To test how a mild and temporal warm-up event can affect the ice abundances of COMs, we used the B~68 Compromise model that undergoes a temperature increase to 20K for 1kyr at $t=1.6$Myr.

\subsubsection{Chemistry differences with previous models}

Figures \ref{att-compr1} and \ref{att-compr2} show that the temperature spike positively affects the production of a number of solid species -- HCOOCH$_3$, CH$_2$CCO, CH$_2$CO, NH$_2$CHO, C$_2$H$_2$, CH$_4$. Notably, the gas phase abundance is substantially increased for species synthesized on the surface -- CH$_2$CCO, H$_2$CO, NH$_2$CN, C$_2$H$_2$, CH$_3$CN, CH$_2$NH, and CH$_4$, thanks to reactive desorption.

For most species, the changes in ice abundance are within one order of magnitude. However, for methyl formate HCOOCH$_3$ and formamide NH$_2$CHO an increase by approximately five and three orders of magnitude is observed, respectively. We therefore explore deeper the temperature spike-induced chemistry of these two species. Initiated at 1.6Myr, the rise of abundances actually occurs over a period of 100kyr after the temperature spike. Figure~\ref{add-sho} shows the abundances of NH$_2$CHO and HCOOCH$_3$ in ice sublayers.

Methyl formate production occurs mainly in the middle sublayer~2. This is unlike most other COMs in the Standard and Compromise models that are synthesized in the outer sublayer~1 (figure~\ref{att-compr-sub}). Sublayer~2 is the only one that is rich in both, CO and H$_2$O. The former is the source species for HCO and CH$_3$O radicals that combine into HCOOCH$_3$. Meanwhile, the photodissociation of water is the principal source of atomic hydrogen, necessary for the hydrogenation sequence $\rm CO \longrightarrow HCO \longrightarrow H_2CO \longrightarrow CH_3O; CH_2OH$. Thus, CO hydrogenation by the mobile H atoms at the 20K spike, followed by combination of HCO and CH$_3$O $\approx$100kyr after the spike, is the main mechanism for the formation of HCOOCH$_3$.

The case of formamide is different -- NH$_2$CHO reaches similar abundance in all three sublayers after the 20K spike. This is because in none of the sublayers its main parent species -- CO and NH$_3$ -- both have a high abundance. While NH$_3$ is mainly concentrated in sublayer~3, CO is abundant only in sublayers 1 and 2.

\subsubsection{Discussion on temperature spike model results}

The ice abundance increases occur because diffusion, proximity, and activation energy barriers are more effectively overcome at 20K. The subsequent reactions temporary reduce of ice abundances for the radical species HCO, CH$_3$O, and CH$_2$OH. The results shown in figures \ref{att-compr1} and \ref{att-compr2} suggest that even such a small and short warm-up period is sufficient to qualitatively reproduce the abundance of methyl formate in ice and that the ice abundances of other organic species do not become disbalanced. The single-point 0D approach used in the model prevents a quantitative comparison between calculation results and observations.

We suggest two likely causes for a (temporal) temperature increase for a portion of ices in a starless or prestellar core. First, the gas parcel can be exposed to the interstellar radiation because of turbulent motions in the core, as suggested by \citet{Boland82} and \citet{Martinell06}. Second, low-velocity transient shocks within the core \citep{Williams84} may also result in mild heating of the gas. Both of these mechanisms also offer a hypothesis for the desorption mechanism -- either photodesorption or collisional sputtering of the weakly-bound, non-polar ice outer layers. These considerations are supported by the turbulent motions observed in starless cores \citep{Redman06,Levshakov14,Steinacker14}.

\section{Summary}
\label{concl}

A model that describes ices in a detailed way was used to investigate ice chemistry in five different examples of starless cores. While the importance of subsurface ice chemistry has been noted in previous studies \citep{Cuppen07,Kalvans10,Garrod13a}, the most general conclusion from the present study is that chemical processes in ice depend strongly on the particular (sub)layer, where the molecules in consideration are located. This arises because of variations in abundances of major ice components (H$_2$O, CO, CO$_2$, N$_2$, NH$_3$) in different layers on the same grains. This conclusion is in line with the results of a prestellar core model \citepalias{Kalvans15b} and a pseudo-time dependent Monte Carlo model by \citep{Chang14}. Single sublayer approach \citep{Kalvans10,Kalvans13,Garrod13a,Belloche14} is not suitable for representing subsurface chemistry of interstellar ices. This is especially so if the synthesis of minor species is studied.

The H$_2$O and CO ices in dark cores are photoprocessed to CO$_2$ via sequence~(\ref{res1}). This sets limits to core lifetimes, because current observations indicate that the CO$_2$:H$_2$O abundance ratio in ice likely is 44\% or less \citep{Boogert11}. The maximum lifetime for a quiescent and stable molecular core with constant physical conditions (Phase~2) inferred this way in the present model is 650kyr (CB~17~S core). It is hard to pinpoint a precise value because of the limitations in the physical model, and uncertainties regarding the turbulence in dark cores. Taking this into account we can state that the model indicates dark core lifetimes of $<$1Myr. The lower CO$_2$:H$_2$O ratio in Taurus \citep{Whittet07} may indicate that these starless cores may be younger than those in most other molecular clouds, observed by \citet{Boogert11} and \citet{Oberg11b}.

The H$_2$O:CO ice abundance ratio cannot be used as reliable indicator of core age because CO resides mostly on the outer surface of the grains. There, it can be rather easily lost to the gas phase or converted into other species in mild heating events. Such processes can be induced by exposure to interstellar photons or low-velocity shocks within the core, as discussed below.

Because of more efficient desorption, the present model predicts shorter and higher abundance peaks during the cloud collapse phase (or ice formation epoch) for a number of species produced in dense gas or on the surface (section~\ref{r-obs}). These include H$_2$O, CO$_2$, NH$_3$, H$_2$O$_2$, H$_2$S, SO, SO$_2$, OCS, HCOOH, H$_2$CO, CH$_2$CO, CH$_4$, O$_2$, H$_2$S$_2$, N$_2$ etc.

We find that O$_2$ and H$_2$O$_2$ are the main oxygen reservoirs in ice, except water and carbon oxides. O$_2$ forms in the gas and on the surface, and is being converted into the H$_2$O$_2$ via photprocessing in subsurface ice. It was suggested that H$_2$O$_2$ and O$_2$H reach their gas-phase abundance peaks during the freeze out epoch, thanks to an active molecular exchange between the gas and the surface. The principal cause of desorption, necessary for this exchange, is irradiation by interstellar photons.

Calculations show that the HNC:HCN ratio in ice can be as high as 1.75 (in CB~27 core model), while it remains close to or below unity in the gas phase. Sulfur in ice is roughly equally divided between reduced (H$_2$S) and oxidized (SO) forms. Subsurface processing of these sulfur species gives rise to OCS, NS, and H$_2$CS, indicating an interesting diversity for the sulfur chemistry.

Investigating the chemistry of COMs, we found that lower activation energies -- more consistent with experimentally detected values -- for hydrogenation reactions of CO and H$_2$CO may help explaining the observations of COMs in the interstellar medium. A mild and temporal warm-up event helps to produce an abundance of methyl formate in ice that is higher and more consistent with gas-phase HCOOCH$_3$ observations. Such an event also substantially increases the ice abundance of formamide.

Desorption by interstellar photons and ice sputtering in grain collisions may transport the COMs to the gas phase. Both of these mechanisms require either turbulence within the cores or external influence. These findings are probably supported by the fact that two of the dense cores with detected COMs -- L1689b \citep{Bacmann12} and B1-b \citep{Marcelino05,Oberg10,Cernicharo12} -- have nearby star-formation regions \citep{Kirk07,Oberg10}, while the third, L1544 \citep{Vastel14}, is known to be turbulent and on the verge of collapse \citep{Caselli02,Tatematsu14}.

\acknowledgments
I acknowledge the support of Ventspils City Council. This research has made use of NASA’s Astrophysics Data System. I thank the anonymous referee for many useful suggestions that improved the paper.

 \footnotesize{
\bibliography{lowmas}
\bibliographystyle{apj}
}

\end{document}